\documentclass[aps,prd,twocolumn,amsmath,10pt,superscriptaddress,floatfix,nofootinbib]{revtex4-1}

\usepackage{epsfig,amssymb,amsfonts,amsmath,mathtools,bm,color,xcolor,graphicx}

\usepackage{hyperref}
\usepackage{multirow}
\hypersetup{pdftex,colorlinks=true,linkcolor=blue,citecolor=blue,menucolor=black,urlcolor=blue,filecolor=blue}

\graphicspath{{Figures/}}

\begin{document}

\title{The relativistic three-body scattering and the $D^0D^{*+}-D^+D^{*0}$ system}
\author{Xu Zhang}\email{zhangxu@itp.ac.cn}
\affiliation{CAS Key Laboratory of Theoretical Physics, Institute of Theoretical Physics,\\
Chinese Academy of Sciences, Beijing 100190, China}

\begin{abstract}
Scattering amplitudes involving three-particle scattering processes are investigated within the isobar approximation which respects constraints from two- and three-body unitarity. The particular system considered is the $D^0D^{*+}-D^+D^{*0}$, where the $D^{*+}$~$(D^{*0})$ enters as a $p$-wave $D^+\pi^0$ or $D^0\pi^+$~($D^0\pi^0$ or $D^+\pi^-$) resonance. The interaction potentials in the coupled-channel $D^0D^{*+}-D^+D^{*0}$ system contain the $\sigma$, $\rho$, $\omega$ and $\pi$-exchange. The analytic continuation of the amplitudes across the three-body unitary cuts
is investigated to search for poles on the unphysical Riemann sheets. Associated with an unstable particle $D^{*+}$~$(D^{*0})$ is a complex
two-body unitarity cut, through which one can further analytically continue into another unphysical Riemann sheet. Dynamical singularities emerged from the $\pi$-exchange potential are stressed. The pole generated from the $D^0D^{*+}-D^+D^{*0}$ interaction and its line shape in
$D^0D^0\pi^+$ break-up production are in agreement with double-charmed tetraquark $T_{cc}^+$ observed by the LHCb Collaboration.
\end{abstract}

\pacs{xxxxxxxxxx}

\maketitle

\section{Introduction}
A large number of new hadron states have been observed experimentally, especially the so-called exotic states which are beyond the scope of the conventional quark model. For a recent review on the experimental and theoretical status of such exotic states, see Refs.~\cite{Hosaka:2016pey,Lebed:2016hpi,Esposito:2016noz,Guo:2017jvc,Olsen:2017bmm,Liu:2019zoy,Brambilla:2019esw,Guo:2019twa,Yang:2020atz,Barabanov:2020jvn,Dong:2021bvy,Chen:2022asf,Mai:2022eur}. Considerable progress has been achieved recently in calculating the hadron spectrum based on first principle 
lattice quantum chromodynamics (QCD) ~\cite{BMW:2008jgk,Dudek:2014qha,Williams:2015cvx,Carrillo-Serrano:2015uca,Eichmann:2016yit,Johnson:2020ilc,Gayer:2021xzv,Lang:2022elg}.
Hadron spectrum emerges from the internal dynamics of the QCD degrees of freedom. The resonance mass characterizes the long-distance dynamics of quarks and gluons, its width manifests on the coupling to the decay channels. To extend our knowledge of this aspect of QCD phenomenology, it is necessary to analytically continue the partial wave amplitudes into the unphysical region and extract resonance parameters from experimental data or lattice QCD simulations, as resonances manifesting as pole singularities in the scattering amplitudes~\cite{Peierls:1959,Newton:1960nws,PhysRev.119.1121,PhysRev.121.1840,PhysRev.123.692,PhysRev.134.B1307,Eden:1964zz,Kato:1965iee,Taylor:1972pty}.

Many of those newly observed exotic states can be interpreted as a deuteron like molecular states generated from various hadron-hadron interactions. 
Among those newly observed exotic states, the most notable example in charmonium sector is the $D^0\bar{D}^{*0}-D^{*0}\bar{D}$ molecule candidate $X(3872)$~\cite{Workman:2022ynf}. However, as discussed in Refs.~\cite{Fleming:2007rp,Dai:2019hrf,Suzuki:2005ha,Baru:2011rs,Schmidt:2018vvl}, the treatment of pions may differ from
ordinary chiral perturbative theory or the $NN$ theory of Refs.~\cite{Kaplan:1998tg,Kaplan:1998we}.
Firstly, since the $\bar{D}^{*0}(D^{*0})-\bar{D}^{0}(D^{0})$ hypefine splitting and the $\pi^0$ mass is nearness, the $\pi^0$-exchange will be characterized by an anomalously small scale and generate anomalously long-range effect. This suggests that if the binding energy of $X(3872)$ is not
much smaller than its upper limit, the $\pi^0$ should be included as explicit degrees of freedom~\cite{Fleming:2007rp,Dai:2019hrf}. Secondly, the $\bar{D}^{*0}$~$(D^{*0})$ is very close to the $\bar{D}^{0}\pi^0$~$(D^{0}\pi^0)$ threshold, and the $\pi^0$ may go on shell in the $D^0\bar{D}^{*0}-D^{*0}\bar{D}$ system. This calls for proper inclusion of the $D^0\bar{D}^0\pi^0$ three-body intermediate state~\cite{Suzuki:2005ha,Baru:2011rs,Schmidt:2018vvl}. Moreover, the inclusion of the $\bar{D}^{*0}$ and $D^{*0}$ finite
widths has a significant effect on the line shape of the $X(3872)$~\cite{Braaten:2007dw,Braaten:2009jke,Hanhart:2010wh}. In Refs.~\cite{Braaten:2015tga,Braaten:2020nmc}, a Galilean-invariant effective field theory for $X(3872)$ is developed to study the interplay between the $D^0\bar{D}^{*0}-D^{*0}\bar{D}$ and $D^0\bar{D}^0\pi^0$ components.
A similar phenomenon happens in the case of newly observed 
double-charm $D^0{D}^{*+}-D^{*+}{D}^0$ molecule candidate $T_{cc}^+$~\cite{LHCb:2021vvq,LHCb:2021auc}.
In Refs.~\cite{Du:2021zzh,Qiu:2023uno}, the effects of the three-body $D^0D^0\pi^+-D^0D^+\pi^0$ dynamics on the pole position of $T_{cc}^+$ and its line shape are studied. In Refs.~\cite{Achasov:2022onn,Dai:2023mxm}, the decay process $T_{cc}^+\to D^0D^0\pi^+$ including the dynamical pion interaction is studied, the contributions from the triangle singularities generated from the $\pi$-exchange in this decay process are discussed in Ref.~\cite{Achasov:2022onn}. Recently, as discussed in Refs.~\cite{Du:2023hlu,Wang:2023iaz,Meng:2023bmz,Hansen:2024ffk}, to correctly extract the pole position of $T_{cc}^+$ from the lattice data~\cite{Padmanath:2022cvl},
the role of left-hand cut contributions from the $\pi$-exchange should be should be properly treated. Thus, to understand the nature of the newly observed exotic states, such as $X(3872)$ and $T_{cc}^+$, it is necessary to 
study the analytic structure of the amplitude involving three-particle scattering.

The aim of this paper is to investigate the relativistic scattering involving three-particle interaction using the isobar approximation~\cite{Blankenbecler:1965gx,Aaron:1968aoz,osti_5335708} which respects constraints from two- and three-body unitarity. 
The specific system considered is the $D^0D^{*+}-D^+D^{*0}$, where the $D^{*+}$~$(D^{*0})$ enters as a $D^+\pi^0$ or $D^0\pi^+$~($D^0\pi^0$ or $D^+\pi^-$) resonance.
We provide a detailed prescription for the analytic continuation of the amplitudes across the three-body and complex two-body unitarity cuts. Dynamical singularities emerged from $\pi$-exchange potential are stressed, and these singularities are inherent to three-body dynamics.

The construction of a unitary $S$-matrix theory for a relativistic three-body scattering has been studied extensively over the past half century~\cite{Blankenbecler:1961zz,Cook:1962zz,Fleming:1964zz,Grisaru:1966uev,Blankenbecler:1965gx,Aaron:1968aoz,Aaron:1973ca,Amado:1974za,osti_5335708} and also in recent investigations~\cite{Mai:2017vot,Jackura:2018xnx,Mikhasenko:2019vhk,Jackura:2019bmu,Dawid:2020uhn}; new insights into covariant vs time-ordered formulations for the interaction kernel were obtained recently~\cite{Zhang:2021hcl}; decay amplitudes with three particles in the final-state may be calculated using Khuri-Treiman equations~\cite{Khuri:1960zz,Aitchison:1966lpz,Pasquier:1968zz,Niecknig:2015ija,Niecknig:2017ylb,Isken:2017dkw,Gasser:2018qtg,Albaladejo:2019huw}, which contains particular realization of constrain from thee-body unitarity as shown in Refs.~\cite{Aitchison:1966lpz,Pasquier:1968zz}. It turns out that the resulting equations are quite involved and demanding to solve.

Compared with the two-particle scattering,
complications arise for the three-particle scattering not only because of the increase in the number of variables necessary to describe the processes, but also the possible appearance of the
dynamic $\pi$-exchange, three-body and complex two-body unitarity cuts as shown in analysis the cut structure of the amplitude within the $S$-matrix theory~\cite{PhysRev.130.2580,Hwa:1964ujn,Rubin:1966zz} and Faddeev equations~\cite{Brayshaw:1968yia,Brayshaw:1969tr,Orlov1984}. Contour deformation of momentum has been employed as a solution tool necessary to analytically continue the amplitude into the unphysical region as shown in tracing the pole trajectory of three-neutron interaction~\cite{Glockle:1978zz} and describing the resonance pole generated from $\pi d$-$N\Delta$ interaction~\cite{Pearce:1984ca,Afnan:1991kb}. We closely follow the approach in Refs.~\cite{Glockle:1978zz,Pearce:1984ca,Afnan:1991kb} and investigate the analytic continuation of the three-body $D^0D^{*+}-D^+D^{*0}$ scattering amplitude. A similar discussion of the analytic continuation involving three-body dynamics can be found in studies of the $\Lambda(\Sigma^-) nn$ interaction~\cite{Matsuyama:1991bm} and $\Lambda d$-$\Sigma d$ interaction~\cite{Afnan:1993pb}, coupled-channel meson-nucleon interaction within the J\"ulich/Bonn/Washington~\cite{Doring:2009yv} and BAC/ANL-Osaka approaches~\cite{Suzuki:2010yn}. Using a relativistic formula for $\pi^+\pi^-\pi^-$ interaction, the decay process $a_1(1260)^-\to \pi^+\pi^-\pi^-$ is studied~\cite{Janssen:1994uf,Sadasivan:2020syi,Sadasivan:2021emk}.  Recently, in Ref.~\cite{Dawid:2023jrj}, a detailed and general prescription for analytic continuation of relativistic three-particle scattering amplitudes is studied, where the subsystem two-body interaction is described in the leading order effective range expansion. 

The paper is organized as follows. In Section~\ref{sec:IntEQ}, we present the integral equation for relativistic $D^0D^{*+}-D^+D^{*0}$ scattering which respects constraints from two- and three-body unitarity. The two-body 
subsystem $D^+\pi^0$~$(D^0\pi^+,D^0\pi^0,D^+\pi^-)$ interacts via a separable interaction. 
In Section~\ref{sec:LagAndPot}, we construct the one-boson-exchange (OBE) potentials, which contain the 
$\sigma$, $\rho$, $\omega$, and $\pi$-exchange. In Section~\ref{sec:AnaCont}, we present the prescription for the analytically continuation of the scattering amplitude into the unphysical region. In Section~\ref{sec:Results}, we present fitting results of the pole position generated from the integral equation and show its line shape in $D^0D^0\pi^+$ final state. In the last section, some conclusions are given.

\section{The basic formalism}\label{sec:IntEQ}

Since the $D^{*0}$ and $D^{*+}$ are unstable and have a width, they can never have an asymptotic state.
In this work, we closely follow the isobar approach, the $D^0D^{*+}-D^+D^{*0}$ system is studied using a relativistic three-body equation, where the $D^{*+}$~$(D^{*0})$ enters as a $D^+\pi^0$ or $D^0\pi^+$~$ (D^0\pi^0$ or $D^+\pi^-)$ resonance.
Assuming only two- and three-body intermediate states, the form of the isobar propagator and the $\pi$-exchange potential are fixed by matching the effective Bethe-Salpeter (BS) equation with unitarity condition as derived in Refs.~\cite{Blankenbecler:1965gx,Aaron:1968aoz,osti_5335708}. \\

\subsection{The coupled-channel $D^0D^{*+}-D^+D^{*0}$ scattering}
We start by constructing the $D^0D^{*+}-D^+D^{*0}$ transition amplitude $T(s,p',p)$ in total angular momentum $J=1$ nonperturbatively.
Within the isobar approach, the $D^0D^{*+}-D^+D^{*0}$ interaction is constructed by an effective BS equation. The partial wave effective BS equation can be written as 
\begin{widetext}
\begin{align}
\label{eq:bse}
{T}(s,p',p)=&{V}(s, p', p)+\int_0^{\Lambda} \frac{k^2dk}{(2\pi)^32\omega(k)}{V}(s, p', k)\tau(\sigma_k){T}(s,k,p),
\end{align}
where 
\begin{align}
\label{eq:poten}
{V}(s,p',p) =
\begin{pmatrix} 
V^{11}_{L'L}(s,p',p) & \  V^{12}_{L'L}(s,p',p)\\       \\  
V^{21}_{L'L}(s,p',p) & \  V^{22}_{L'L}(s,p',p)
\end{pmatrix}, \qquad
\tau(\sigma_k) = 
\begin{pmatrix} 
                           \tau^1(\sigma_k)&  \   0             \\          \\  
                              0         &  \  \tau^2(\sigma_k)
\end{pmatrix}, 
\end{align}
\end{widetext}
where, in each matrix element $V^{i'i}_{L'L}(s,p',p)$, the index $i(i')=1,2$ labels the particle channel ($D^0D^{*+}=1$, $D^+D^{*0}=2$) and $L(L')$ denotes the orbital angular momentum. The same structure holds for ${T}(s,p',p)$. The isobar propagator will be given in Section~\ref{sec:IsoPro}. The particles in channel 1 are
\begin{eqnarray}
1={D^{*+}},\quad 2={D^{0}}, 
\end{eqnarray}
and in channel 2 are 
\begin{eqnarray}
 1={D^{*0}},\quad 2={D^{+}}.
\end{eqnarray}
The energy $\omega(k)$ is $\omega_{i,2}(k)$ in channel $i$, where $\omega_{i,j}(k)=\sqrt{m_{i,j}^2+k^2}$.
The incoming and outgoing momenta are denoted as $p$ and $p'$, $s$ denotes the squared invariant mass of the three-body system, and $\sigma_k$ denotes the square of the four-momentum of the intermediate $D^{*+}$ or $D^{*0}$ subsystem, 
\begin{eqnarray}
\label{eq:TwoInv}
\sigma_k=s-2\sqrt{s}\,\omega_{i,2}(k)+m_{i,2}^2.
\end{eqnarray}
We take the masses of the mesons to be~\cite{Workman:2022ynf}
\begin{eqnarray}
\nonumber &&m_{D^0}=1.86484\,\,{\rm GeV},\quad m_{D^{*+}}=2.01026 \,\,{\rm GeV}, \\
&&m_{D^+}=1.86966\,\, {\rm GeV},\quad m_{D^{*0}}=2.00685\,\, {\rm GeV}.
\end{eqnarray}
The integrations in Eq.~\eqref{eq:bse} and Eq.~\eqref{eq:decayamp} will be regularized by the same cutoff $\Lambda$ in contrast to a form factor, 
since it simplifies the analytic continuation of the effective BS equation as discussed in Section~\ref{sec:AnaCont}.

To obtain the the partial wave interaction potentials in the $JLS$ basis, we use the method given in Refs.~\cite{Jacob:1959at,Chung:1971ri}. Firstly, in the helicity basis the relevant partial wave is extracted. We choose the incident $\vec{p}$ along the $z$-axis and outgoing $\vec{p'}$ to be in the $x,z$ plane.
In helicity basis, the relevant partial wave is extracted exploiting the orthonormality of Wigner $d_{\lambda\lambda'}^J(\rm{cos}\theta)$ functions,
\begin{eqnarray}
\label{eq:patwa}
V_{\lambda'\lambda}^J(s,p',p)=2\pi\int_{-1}^{+1}d_{\lambda\lambda'}^J({\rm cos}\theta)V_{\lambda'\lambda}(s,p',p) \,d \rm{cos}\theta,
\end{eqnarray}
where $\lambda(\lambda')=0,\pm 1$ is the helicity eigenvalue of the spin-1 meson $D^{*+}$ or $D^{*0}$.
Then, the transition from the helicity to the $JLS$ representation is given by
\begin{eqnarray}
 V^{J}_{L'L}(s,p',p)=\sum_{\lambda'\lambda}\langle JL'S |J\lambda'\rangle V_{\lambda'\lambda}^J(s,p',p) \langle J \lambda|JLS\rangle,
\end{eqnarray}
where 
\begin{eqnarray}
 U_{L\lambda}=\langle JLS |J\lambda\rangle=\sqrt{\frac{2L+1}{2J+1}} \langle L0S\lambda|J\lambda\rangle \langle S \lambda 0 0 | S \lambda\rangle.
\end{eqnarray}
Here, $J$ and $S$ denote the total angular momentum and $D^{*+}$ or $D^{*0}$ spin, respectively. The symbols on the right-hand side are standard $SU(2)$ Clebsch-Gordan coefficients. In the present work, we will only consider $L=L'=0$. Then for $J=S=1$, $L=0$ and $\lambda=0,\pm 1$, we get
\begin{eqnarray}
\langle JLS |J\lambda\rangle=\frac{1}{\sqrt{3}}.
\end{eqnarray}

\subsection{The isobar propagator}
\label{sec:IsoPro}
\begin{figure}[tbhp]
\begin{center}
\includegraphics [scale=0.38] {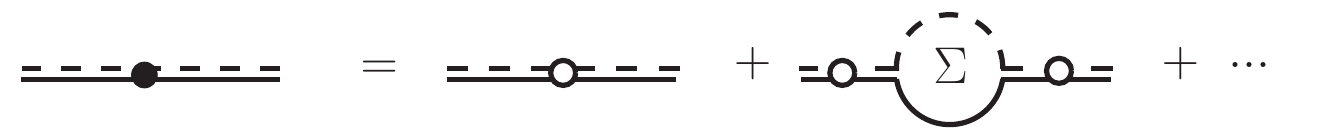}
\caption{A graphical representation of the isobar propagator of Eq.~\eqref{Eq:IsobarA}. The isobar propagator ${D^{*+}}$ (${D^{*0}}$) is dressed by an infinite number of $D^0\pi^+$ and $D^+\pi^0$ ($D^0\pi^0$ and $D^+\pi^-$) bubbles.
The dashed and solid lines denote the $D^{0,+}$ and $\pi^{\pm,0}$, respectively.
}
\label{fig:SelEnG}
\end{center}
\end{figure}

The isobar propagator for ${D^{*+}}$ and ${D^{*0}}$ is shown in Fig.~\ref{fig:SelEnG}.
For the present work, we choose the renormalized isobar propagator 
\begin{eqnarray}
\label{Eq:IsobarA}
\nonumber && \!\!\!\!\!\!\!\!\!\!\!\!\!\!\! \tau^{i}(\sigma_k)=  \\
&&\!\!\!\!\!\!\!\!\!\!\!\!\!\!\frac{1}{\sigma_k-m_{i,1}^2-\Sigma_{i,34}^R(\sigma_k)-\Sigma_{i,3'4'}^R(\sigma_k)+im_{i,1}\Gamma_{i,\gamma}+i\epsilon}, 
\end{eqnarray}
where 
\begin{eqnarray}
\Sigma_{i,jk}^R(\sigma_k)&=&\Sigma_{i,jk}(\sigma_k)-{\rm Re} \Sigma_{i,jk}^{\rm sub}(m_{i,1}^2), 
\end{eqnarray}
and  
\begin{widetext}
\begin{eqnarray}
\label{eq:sefa}
\Sigma_{i,jk}^{\rm sub}(m_{i,1}^2) 
=\Sigma_{i,jk}(m_{i,1}^2) 
+\Big(\sigma_k-m_{i,1}^2 \Big)\Big(\frac{d}{d{\sigma_k}} \Sigma_{i,jk}(\sigma_k) |_{\sigma_k=m_{i,1}^2}\Big).
\end{eqnarray}
$\Gamma_{i,\gamma}$ is the radiative decay width of $D^{*+}$ or $D^{*0}$.
Since the radiative decay $D^{*0}\to D^0\gamma$ has a large branching ratio $\mathcal{B}\equiv \Gamma_{D^{*0},\gamma}/\Gamma_{D^{*0},\rm total}=35.3\%$~\cite{Workman:2022ynf}, we have included the radiative decay widths of $D^{*+}$ and $D^{*0}$. 
To obtain the renormalized isobar propagator $\tau^{i}(\sigma_k)$ in
Eq.~\eqref{Eq:IsobarA}, twice subtraction at the $D^{*+}$ or $D^{*0}$ physical mass has been used. The self-energy $\Sigma_{i,jk}(\sigma_k)$ is
\end{widetext}
\begin{eqnarray}
\label{eq:sefbar}
\nonumber&&\!\!\!\!\!\!\!\!\!\! \Sigma_{i,jk}(\sigma_k)= \\
&&\!\!\!\!\!\!\!\int_0^{\infty}\frac{ l^2dl}{(2\pi)^3}\frac{\omega_{i,jk}(l)}{2{\omega_{i,j}(l)\omega_{i,k}(l)}} \frac{}{} \frac{v_{ijk}(l)^2}{\Big(\sigma_k-\omega_{i,jk}(l)^2+i\epsilon\Big)}, \quad
\end{eqnarray}
where
\begin{eqnarray}
\omega_{i,jk}(l)=\omega_{i,j}(l)+\omega_{i,k}(l),
\end{eqnarray}
where $j(k)=3(4)$ or $3'(4')$ labels a particle, and the particles in channel 1 are
\begin{eqnarray}
 3=D^0, \quad 4=\pi^+, \quad
 3'=D^+, \quad 4'=\pi^0,
\end{eqnarray}
and in channel 2 are 
\begin{eqnarray}
 3=D^0, \quad 4=\pi^0, \quad
 3'=D^+, \quad 4'=\pi^-.
\end{eqnarray}
The vertex $v_{ijk}(l)$
in the self-energy will be given in Eq.~\eqref{Eq:TwoInpu} in Section~\ref{sec:TwoBody}.
The self-energy $\Sigma_{i,jk}(\sigma_k)$ contains the two-body intermediate state, and reflects the multi-Riemann sheet structure of the scattering amplitude. 
The prescription for the analytically continuation of the scattering amplitude into the unphysical region is present in Section~\ref{sec:AnaCont}.
\begin{figure*}[tbhp]
\begin{center}
\includegraphics [scale=0.37] {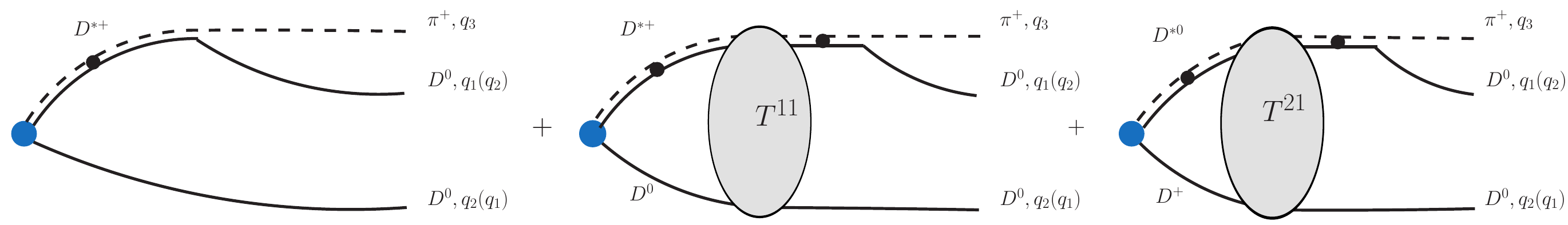}
\caption{Graphical representation for $D^0D^0\pi^+$ production. The short-range dynamics corresponding to $T_{cc}^+$ production can be absorbed to an overall coefficient $\mathcal{F}$. 
}
\label{fig:DecDM}
\end{center}
\end{figure*}
\subsection{Three-body break-up production process in $D^0D^0\pi^+$ final state}
Now, we calculate the three-body $D^0D^0\pi^+$ production rate using the constructed scattering amplitude $T(s,p',p)$. This quantity was measured by the LHCb Collaboration~\cite{LHCb:2021vvq,LHCb:2021auc} and thus serves as an important link between theory and experiment. We consider a process
in which $T_{cc}^+$ production at short ranges and subsequently decays
to $D^0D^0\pi^+$. And as point out in Ref.~\cite{Braaten:2007dw}, the short-range dynamics corresponding to
$T_{cc}^+$ production can be absorbed into an overall coefficient $\mathcal{F}$. 

In Fig.~\ref{fig:DecDM}, we show a graphical representation for the $T_{cc}^+$ decay process. The production amplitude is separated into connected and disconnected parts using the LSZ reduction~\cite{Itzykson:1980rh,Lehmann:1954rq}.
The decay amplitude $\mathcal{M}_{\Lambda\lambda}$ describes the $T_{cc}^+$ resonance at rest with helicity $\Lambda$ measured along $z$ axis into a $D^0$ and a $D^{*+}\to D^0\pi^+$ with helicity $\lambda$ and can be written as ~\cite{Jacob:1959at,Chung:1971ri,Sadasivan:2020syi,Sadasivan:2021emk}
\begin{eqnarray}
\nonumber \mathcal{M}_{\Lambda\lambda}(\vec{q}_1,\vec{q}_2,\vec{q}_3)&=&\frac{\mathcal{F}}{\sqrt{2}}\Big[\sqrt{\frac{3}{4\pi}}\mathcal{D}_{\Lambda\lambda}^{1*}(\phi_1,\theta_1,0)\\
\nonumber&&\mathcal{M}_L(q_1)U_{L\lambda}v_{\lambda}(\vec{q}_2,\vec{q}_3)+(\vec{q}_1 \leftrightarrow \vec{q}_2)\Big],\\
\end{eqnarray}
and 
\begin{eqnarray}
\label{eq:decayamp}
\nonumber &&\mathcal{M}_L(q_1)=\Big( g_L^1 \\
\nonumber &&+\int_0^{\Lambda}\frac{p^2dp}{(2\pi)^32\omega_{D^{0}}(p)}g_{L'}^1\tau^1(\sigma_p)T_{L'L}^{11}(p,q_1)\\
\nonumber &&+\int_0^{\Lambda}\frac{p^2dp}{(2\pi)^32\omega_{D^{+}}(p)}g_{L'}^2\tau^2(\sigma_p)T_{L'L}^{21}(p,q_1)\Big) \tau^1(\sigma_{q_1}),\\
\end{eqnarray}
where $\vec{q}_1$, and $\vec{q}_2$ are the outgoing $D^0$ momentum, and $\vec{q}_3$ is the outgoing $\pi^+$ momentum. $\mathcal{D}_{\Lambda\lambda}^{1*}(\phi_1,\theta_1,0)$ denotes the capital Wigner-$D$ function, with $\theta_1$ and $\phi_1$ giving the polar and azimuthal angles of $\vec{q}_1$, respectively. The vertex $v_{\lambda}(\vec{q}_2,\vec{q}_3)=-ig_{DD^*\pi}\epsilon^{\mu}_{\lambda}(\vec{q}_2+\vec{q}_3) q_{3\mu}$ is given from the Lagrangian in Appendix~\ref{sec:APPLa}. 
The polarization vectors of the spin-1 particles are given in Appendix~\ref{sec:Polari}.
In addition, $g_{L}^i$ is the effective coupling of $T_{cc}^+$ to channel $i$. 
In the exact isospin limit, one would have $g_{L}^{1}=-g_{L}^{2}$ for an isoscalar state.

The production rate is given by a phase space integral over the decay amplitude, and can be written as 
\begin{eqnarray}
\label{eq:lineshape}
\nonumber \frac{d\Gamma(\sqrt{s})}{d\sqrt{s}} &=&\int  \frac{1}{(2\pi)^5}\frac{1}{16{s}}
 \, \Big(\frac{1}{3}\sum_{\Lambda}|\sum_{\lambda}\mathcal{M}_{\Lambda\lambda}(\vec{q}_1,\vec{q}_2,\vec{q}_3)|^2\Big) \\
 &&\times q_3^*\,q_1\,dm_{23}d\Omega_3^*d\Omega_1,
\end{eqnarray}
where $q_3^*$ and $\Omega_3^* (\theta_3^*,\phi_3^*)$ are the momentum and angle of particle 3 in the rest frame of particles 2 and 3,
$q_1$ and $\Omega_1(\theta_1,\phi_1)$ are the momentum and angle of particle 1 in the rest frame of the decaying particle. The momentum $q_3^*$ and $q_1$ are
\begin{eqnarray}
 q_3^*=\frac{\lambda^{1/2}(\, m_{23}^2, m_2^2,m_3^2\,)}{2m_{23}},\quad q_1=\frac{\lambda^{1/2}(\,s,m_{23}^2,m_1^2\,)}{2\sqrt{s}},
\end{eqnarray}
where
\begin{eqnarray}
\lambda(x^2,y^2,z^2)=[x^2-(y+z)^2][x^2-(y-z)^2].
\end{eqnarray}
When one exploits the azimuthal symmetry, the integral variable $\phi_1$ is trivial. In our calculation, we integrate over four variables: $m_{23}$, $\theta_1$, $\theta_3^*$ and $\phi_3^*$.

\section{The interaction potential}\label{sec:LagAndPot}
In the present work,  a crucial input is the interaction potentials between the 
charm mesons. The interaction potentials, which enter the effective BS equation, contain 
the $\sigma$, $\rho$ and $\omega$ and $\pi$-exchange. The interaction potentials will be constructed with the help of effective Lagrangian given in Appendix~\ref{sec:APPLa}.
For the $\sigma$, $\rho$ and $\omega$-exchange, the interaction potentials will be constructed in 
a covariant form. For the $\pi$-exchange, the form of the interaction potential is derived from dispersive techniques as shown in Refs.~\cite{Blankenbecler:1965gx,Aaron:1968aoz,osti_5335708}.
Moreover, with the help of the effective Lagrangian,
the two-body input $v_{ijk}(l)$ in the self-energy is given in Eq.~\eqref{Eq:TwoInpu}.
\subsection{Two-body subsystem interaction}\label{sec:TwoBody}
The fundamental ingredient of the three-body theory developed in Section~\ref{sec:IntEQ} is the assumption of an underlying separable two-body interaction. This implies that the two-body $D^+\pi^0$ or $D^0\pi^+$ ($D^0\pi^0$ or $D^+\pi^-$) scattering amplitude in total angular momentum $J_{\rm sub}=1$ channel is generated exclusively by a $D^{*+} (D^{*0})$ pole diagram. 
To obtain the partial wave projected vertex $v_{ijk}(p)$ in the self-energy in Eq.~\eqref{eq:sefbar}, we can consider the first-order Born series for the $j(p_j)k(p_k) \to j(p_j')k(p_k')$ scattering in two-body rest frame $p_j^{(\prime)} + p_k^{(\prime)}=(\sqrt{\sigma},0)$. This amplitude can be written as 
\begin{eqnarray}
\label{Eq:TwoBody}
\nonumber {\cal A}(\sigma, z)&=&I_f\cdot g_{\rm DD^{*}\pi}^2 \frac{\sum_{\lambda} \epsilon_{\lambda,\mu}
(p_j+p_k) p_{k}^{\mu} \epsilon_{\lambda,\nu}^{*}(p_j'+p_k') p_{k}^{\nu \prime}}{\sigma-m_{i,1}^2+i\epsilon}\\
&=&I_f\cdot g_{\rm DD^{*}\pi}^2 \frac{pp'z}{\sigma-m_{i,1}^2+i\epsilon},
\end{eqnarray}
where $z=\vec{p}\cdot\vec{p'}/pp'$, $p_j=(\,w_{i,j}(p),\,\vec{p}\,)$, and analogous variables for $p_k$, $p_j'$ and $p_k'$.
$I_{f}$ denotes the isospin factor. For $D^0\pi^+ \to D^0\pi^+ $ and $D^+\pi^0\to D^+\pi^0$ interactions, $I_f$ are 1 and ${\frac{1}{2}}$, respectively. For $D^0\pi^0\to D^0\pi^0$ and $D^+\pi^-\to D^+\pi^-$ interactions,
$I_f$ are ${\frac{1}{2}}$ and 1, respectively. The second equal sign in Eq.~\eqref{Eq:TwoBody} is due to the helicity sum.
Projecting this amplitude to the $P$-wave amounts then to 
\begin{eqnarray}
\label{Eq:TwoInpu}
\nonumber {\cal{A}}^1(\sigma)&=&2\pi \int_{-1}^{1} dz P_1(z) {\cal{A}}(\sigma,z) \\
\nonumber &=& \frac{\Big(g_{\rm DD^{*}\pi}\sqrt{\frac{4\pi}{3}I_f}\cdot {p}\Big) \Big(g_{\rm DD^{*}\pi}\sqrt{\frac{4\pi}{3}I_f}\cdot {p'}\Big)} {\sigma-m_{i,1}^2+i\epsilon}\\
&=&\frac{v_{ijk}(p)v_{ijk}(p')}{\sigma-m_{i,1}^2+i\epsilon},
\end{eqnarray}
and we obtain the projected vertex $v_{ijk}(p)$.

The two-body dynamics are encoded in $v_{ijk}(p)$, but also in the self-energy $\Sigma_{i,jk}(\sigma_k)$. 
The full two-body $j(p_j)k(p_k) \to j(p_j')k(p_k')$ scattering amplitude can be written as
\begin{eqnarray}
\label{Eq:TwoSub}
\nonumber{\cal A}(\sigma)= \frac{v_{ijk}(p)v_{ijk}(p')}{\sigma-m_{i,1}^2-\Sigma_{i,34}^R(\sigma)-\Sigma_{i,3'4'}^R(\sigma)+im_{i,1}\Gamma_{i,\gamma}+i\epsilon}, \\
\end{eqnarray}
the renormalized self-energy is given in Eq.~\eqref{eq:sefa}. It is obvious that in the three-body system,
the two-body subsystem invariant mass is obtained by replacing $\sigma \to \sigma_k$, by the inclusion of the spectator $D^0$ or $D^+$.\\

\begin{table}[hbtp]
\caption{The isospin factors $IF$ for Type-A and Type-B diagrams and different exchange mesons.}
\label{Tab:IsoSpin}
\begin{ruledtabular}
\setlength{\tabcolsep}{1.8mm}{
\begin{tabular}{c|ccc|ccccc}
&\multicolumn{3}{c|}{Type-A} &\multicolumn{5}{l}{ \ \ \ \ \ \ \ \ \ \ \ Type-B}\\
           &$\rho^0$       &$\rho^{\pm}$&$\omega$      &$\rho^0$  &$\rho^{\pm}$ &$\omega$ &$\pi^0$ & $\pi^{\pm}$ \\ \hline
$1\to 1$   & {\multirow{2}{*}{$\frac{1}{2}$}} &{\multirow{2}{*}{ --}}         & \multirow{2}{*}{$-\frac{1}{2}$}  & {\multirow{2}{*}{-- }}           &{\multirow{2}{*}{ $1$}}          & {\multirow{2}{*}{ -- }}    &{\multirow{2}{*}{ -- }}    & {\multirow{2}{*}{$1$}} \\ 
$2\to 2$  &&&&&& \\ \hline
$1\to 2$   & --            & $-1$       &         --      & $-\frac{1}{2}$ & --      &  $\frac{1}{2}$     & $-\frac{1}{2}$ & --\\
\end{tabular}}
\end{ruledtabular}
\end{table}
\subsection{The interaction potential}
\begin{figure*}[tbhp]
\begin{center}
\includegraphics [scale=0.4] {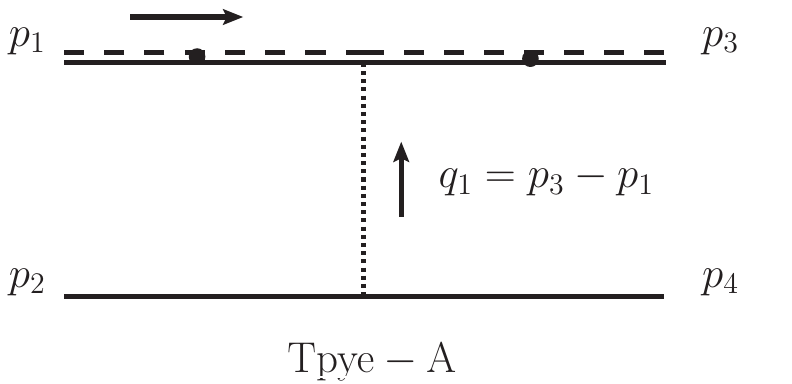} \quad\quad \quad
\includegraphics [scale=0.4] {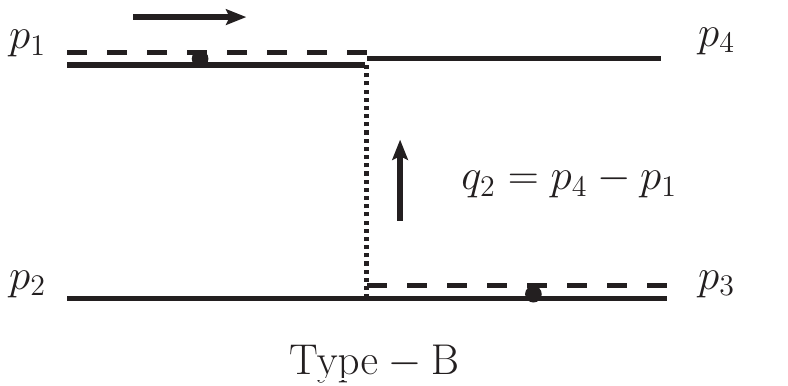}
\caption{Feynman diagrams for meson exchange between the charm mesons. The notations are the same as those in Fig.~\ref{fig:SelEnG}.
The arrows denote the direction of momenta.
}
\label{fig:OnePEch}
\end{center}
\end{figure*}
In this work, we consider two types of Feynman diagrams for meson exchange between the charm mesons as shown in Fig.~\ref{fig:OnePEch}.
Then for coupled-channel $D^0D^{*+}-D^+D^{*0}$ interaction,
with the Lagrangians given in Appendix~\ref{sec:APPLa}, the interaction potentials can be written as 
\begin{widetext}
\begin{eqnarray}
\nonumber \langle p'\lambda'|{V}_{A-V}(E)| p\lambda\rangle&=&-g_{{\rm DD}V}\cdot g_{{\rm D^*D^*}V}\cdot IF\cdot
(p_3+p_1)^{\mu}\frac{-g_{\mu\nu}+q_{1\mu}q_{1\nu}/m_V^2}{q_1^2-m_{V}^2+i\epsilon}(p_4+p_2)^{\nu}\epsilon_{\lambda'}^{*\alpha}(p_3)\epsilon_{\lambda \alpha}(p_1)\\
&&+2g_{{\rm DD}V}\cdot g_{{\rm D^*D^*}V}'\cdot IF\cdot[\epsilon_{\lambda'}^{*\mu}(p_3) \epsilon_{\lambda}^{\alpha}(p_1) q_{1\alpha}-
\epsilon_{\lambda'}^{*\alpha}(p_3) \epsilon_{\lambda}^{\mu}(p_1) q_{1\alpha}]
\frac{-g_{\mu\nu}}{q_1^2-m_{V}^2+i\epsilon}(p_4+p_2)^{\nu},\\
\langle p'\lambda'|{V}_{B-V}(E)| p\lambda\rangle&=& g_{{\rm DD^*}V}^2\cdot IF\cdot \varepsilon_{\alpha'\beta'\mu'\nu'} (p_3+p_2)^{\alpha'} q_2 ^{\beta'}\epsilon_{\lambda'}^{*\nu'}(p_3) \frac{-g_{\mu' \mu}}{q_2^2-m_{V}^2+i\epsilon}  \varepsilon_{\alpha\beta\mu\nu} (p_4+p_1)^{\alpha} q_2 ^{\beta}\epsilon_{\lambda}^{\nu}(p_1), \\  
\label{Eq:OPE} \langle p'\lambda'|{V}_{B-P}(E)| p\lambda\rangle&=& g_{\rm
DD^{*}P}^2\cdot IF\cdot\epsilon_{\lambda'}^{*\nu}(p_3) q_{2\nu}\frac{\omega_{i,2}(p)+\omega_{i',2}(p')+\omega_{P}(q)}{\omega_{P}(q)[s-(\omega_{i,2}(p)+\omega_{i,2'}(p')+\omega_{P}(q))^2+i\epsilon]}  \epsilon_{\lambda}^{\mu}(p_1) q_{2\mu}, \\
\langle p'\lambda'|{V}_{A-S}(E)| p\lambda\rangle&=& g_{\rm D^{*}D^{*}\sigma} g_{\rm DD\sigma} \epsilon_{\lambda'}^{*\mu}(p_3) \epsilon_{\lambda\mu}(p_1)\frac{1}{q_1^2-m_{S}^2+i\epsilon}\,, 
\end{eqnarray}
\end{widetext}
with $\omega_{P}(q)=\sqrt{p^2+p'^2+m_P^2+2pp'{\rm cos}\,\theta}$, and $V$, $P$ and $S$ denote the exchanged vector $\rho^{0,\pm}$, pseudoscalar $\pi^{0,\pm}$ and scalar $\sigma$ mesons, respectively. The four-momentum of the incoming and outgoing particles are $p_1=(\,\sqrt{s}-\omega_{i,2}(p),\,\vec{p}\,)$, $p_2=(\,\omega_{i,2}(p),\,-\vec{p}\,)$, $p_3=(\,\sqrt{s}-\omega_{i',2}(p'),\,\vec{p'}\,)$, and $p_4=(\,\omega_{i',2}(p'),\,-\vec{p'}\,)$. The four-momentum of the exchanged mesons are $q_1=p_3-p_1$ for Type-A diagram, and $q_2=p_4-p_1$ for Type-B diagram.
The isospin factors $IF$ for Type-A and Type-B diagrams are given in Table~\ref{Tab:IsoSpin}. We can see a cancellation between the neutral $\rho^0$ and $\omega$ exchange potentials in both type-A 
and type-B diagrams. And only the charged $\rho^{\pm}$ and $\pi^{0,\pm}$-exchange potentials contribute. Then we build $DD^*$ isoscalar $I=0$ and isovector $I=1$ combinations as
\begin{eqnarray}
\nonumber &&|DD^*,I=0\rangle=-\frac{1}{\sqrt{2}}(D^{*+}D^0-D^{*0}D^+),\\
&&|DD^*,I=1\rangle=-\frac{1}{\sqrt{2}}(D^{*+}D^0+D^{*0}D^+).
\end{eqnarray}
\section{Analytic properties of the integral equation}\label{sec:AnaCont}
Contour deformation of momentum has been employed as a solution tool necessary to analytically continue the amplitude into the unphysical region as shown in tracing the pole trajectory of three-neutron interaction~\cite{Glockle:1978zz} and describing the resonance pole generated from $\pi d$-$N\Delta$ interaction~\cite{Pearce:1984ca,Afnan:1991kb}.
Via complex momentum contour deformation, the $\pi$-exchange cut can be avoided and the analytic domain of the effective BS equation kernel is extended. We closely follow the approach in Refs.~\cite{Glockle:1978zz,Pearce:1984ca,Afnan:1991kb} and investigate the analytic continuation of the three-body $D^0D^{*+}-D^+D^{*0}$ scattering amplitude. In Refs.~\cite{Janssen:1994uf,Sadasivan:2020syi,Sadasivan:2021emk}, some aspects of this approach
were also discussed in detail.

Two types of integrations occur: (i) in $k:=|\vec{k}\,|$ within the effective BS equation in Eq.~\eqref{eq:bse} and (ii) in $l:=|\vec{l}\,|$ within the self-energy term of the two-body subsystem in Eq.~\eqref{Eq:IsobarA}. The corresponding integral contours can be chosen individually and are refereed to in the following as as ``spectator
momentum contour'' (denoted by $\Gamma_{\rm spe}$ ) and ``self-energy contour'' (denoted by $\Gamma_{\rm sef}$), respectively. Both contours start at the respective
origins, $k=l=0$, and end at $k=\Lambda$ and $l=\infty$, respectively. In between these limits,
different choices of the contours define different Riemann sheets in $\sqrt{s}$ as talked in the following.
 
\subsection{Analytical continuation of the self-energy}
\label{sec:AnaSef}
\begin{figure}[tbhp]
\begin{center}
\includegraphics [scale=0.35] {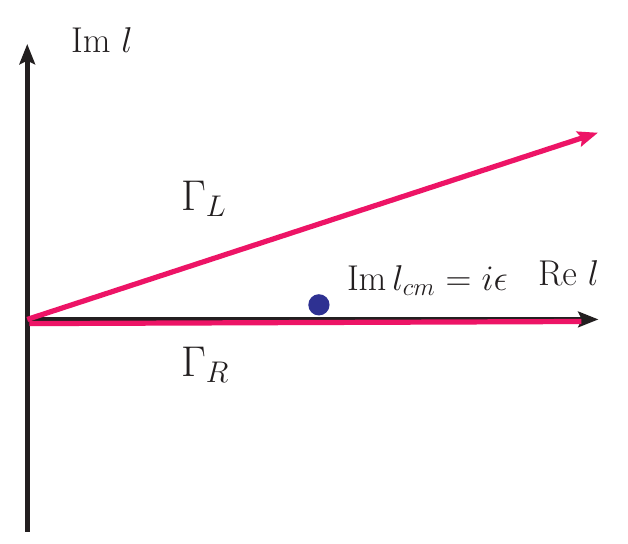}\quad \quad
\includegraphics [scale=0.35] {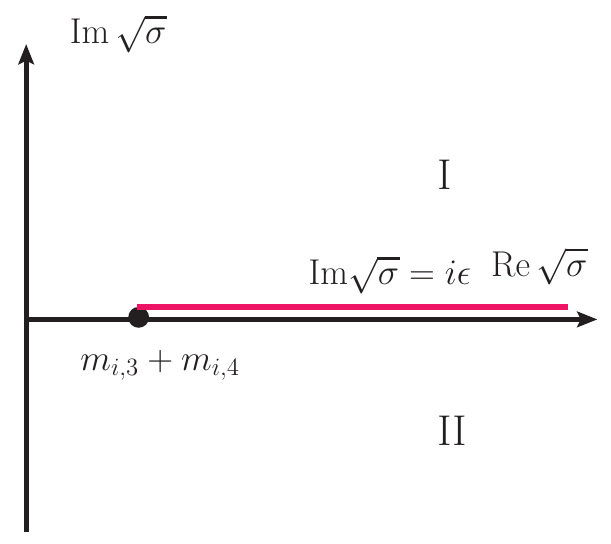}
\includegraphics [scale=0.35] {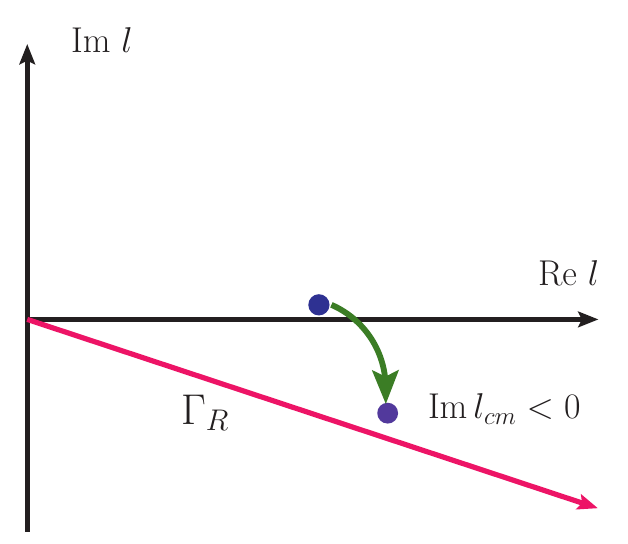}\quad\quad
\includegraphics [scale=0.35] {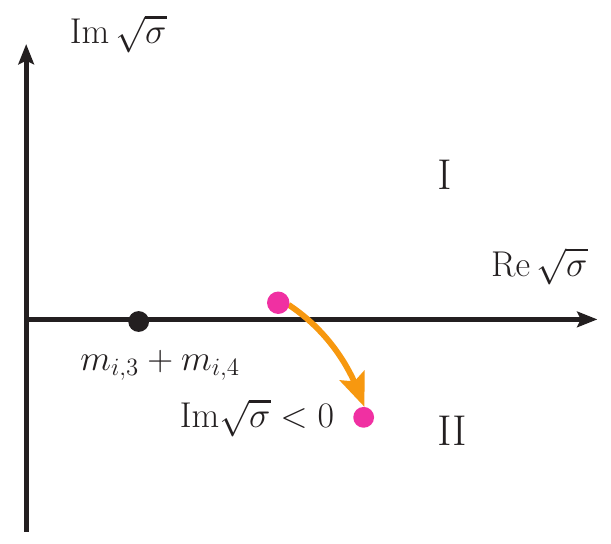}
\caption{Integration contours for the sel-energy in Eq.~\eqref{eq:sefbar}. Upper row:
momentum integral contour (left panel) and the physical Riemann sheet of the self-energy at $\sqrt{\sigma}+i\epsilon$ (right panel). Lower row:
integral contour along the red line (left panel) and the unphysical Riemann sheet of the self-energy at ${\rm Im }\sqrt{\sigma}<0$ (right panel).
}
\label{fig:TBCont}
\end{center}
\end{figure}
As a function of energy $\sqrt{\sigma}$, the self-energy $\Sigma_{i,jk}(\sigma)$ in Eq.~\eqref{eq:sefbar} is no longer single valued. We must therefore
place the complex $\sqrt{\sigma}$ plane by a Riemann surface of multi-sheet,  the different Riemann sheets are connected by a branch cut. A branch point defines the thresholds, and at which a cut begins is fixed. This is illustrated in Fig.~\ref{fig:TBCont} above threshold at $\sqrt{\sigma}=m_{i,3}+m_{i,4}$. The placement of contour $\Gamma_{\rm sef}$ producing physical amplitude is constrained by the $+i\epsilon$ in Eq.~\eqref{eq:sefbar}. In the figure, choosing the integral contour passing the
singularity at $l_{cm}+i\epsilon$ on the right ($\Gamma_{R}$),
\begin{eqnarray}
l_{cm}= \frac{\lambda^{1/2}(\sigma,m_{i,3}^2,m_{i,4}^2)}{2\sqrt{\sigma}},
\end{eqnarray}
yields a self-energy on the physical Riemann sheet. In contrast, choosing the integral contour 
passing the singularity at $l_{cm}+i\epsilon$ on the left ($\Gamma_{L}$) leads to a sign change in ${\rm Im} \Sigma_{i,jk}(\sigma)$ and a self-energy on the unphysical Riemann sheet.
\begin{figure*}[tbhp]
\begin{center}
\includegraphics [scale=0.42] {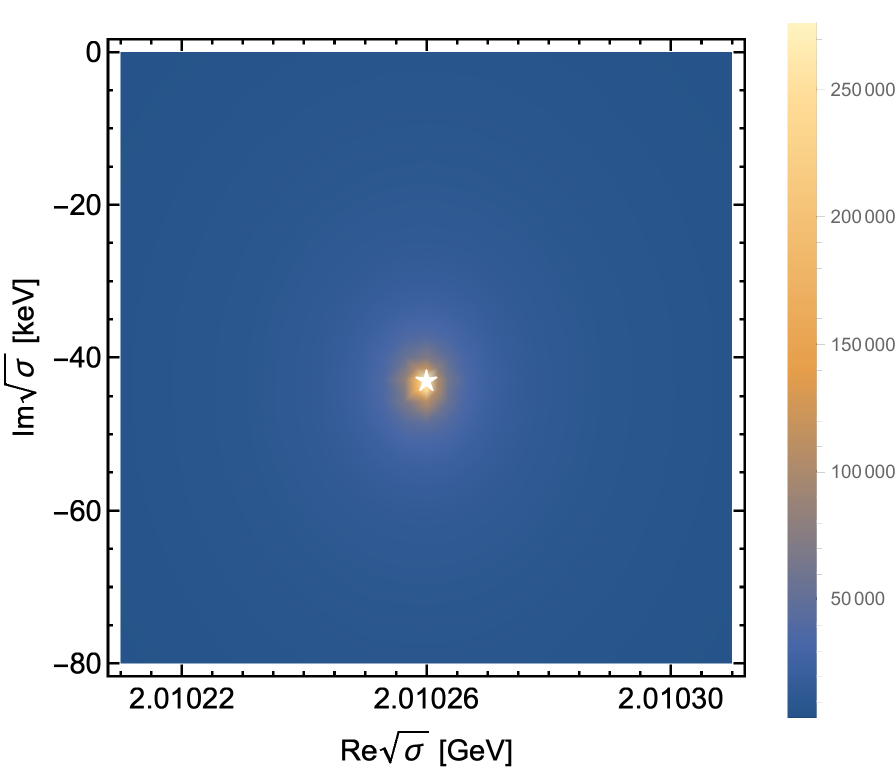}\quad \quad
\includegraphics [scale=0.42] {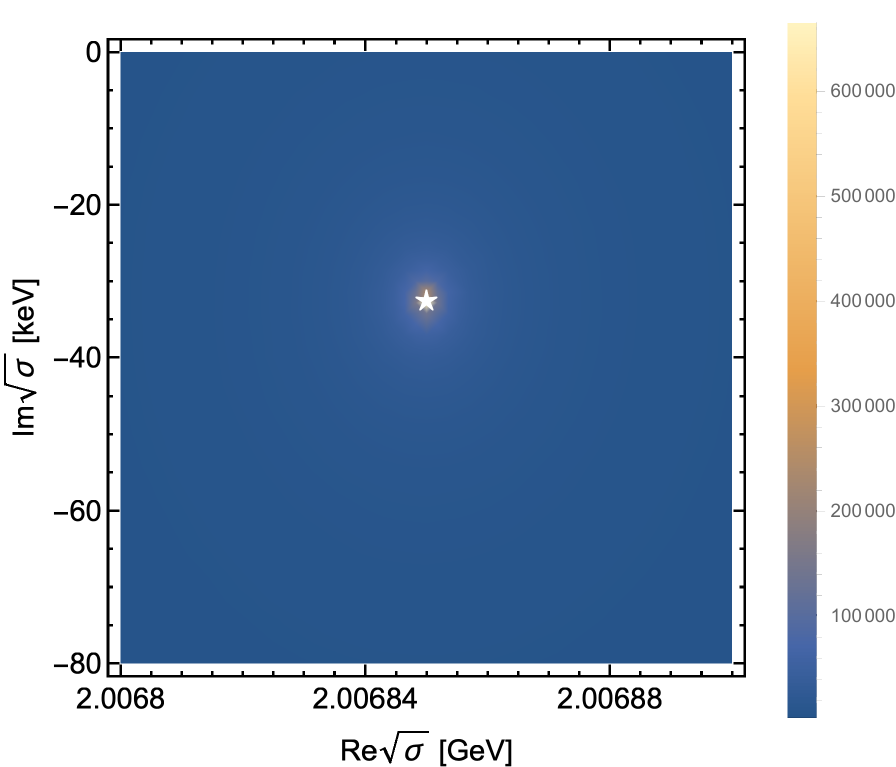}
\includegraphics [scale=0.42] {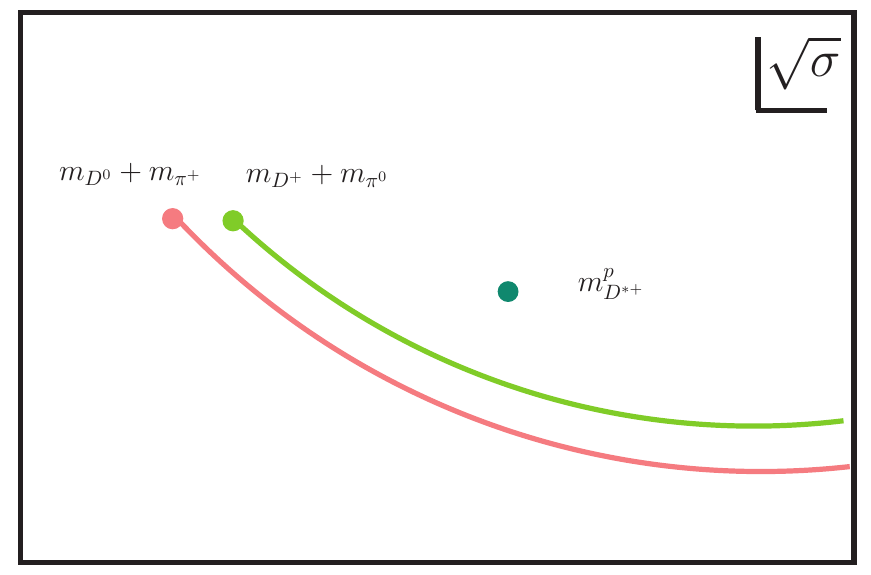}\quad \quad
\includegraphics [scale=0.42] {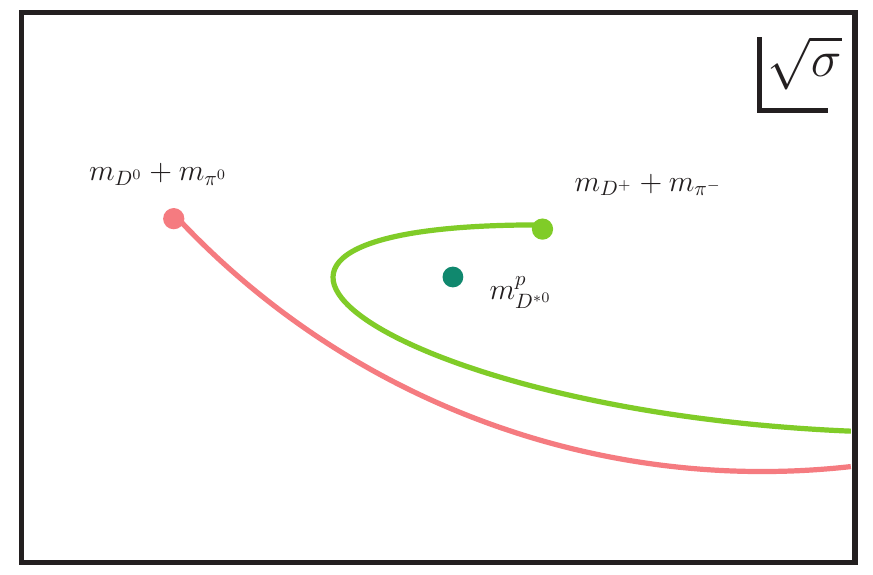}
\caption{Upper row: As the $D^0\pi^+$ and $D^+\pi^0$ ($D^0\pi^0$ and $D^+\pi^-$) channels are on their unphysical Riemann sheets, the pole position of $D^{*+}$ ($D^{*0}$) at the $\sqrt{\sigma}$ plane. The $D^{*+}$ ($D^{*0}$) pole at $\sqrt{\sigma}$ plane is highlighted with the white star in the left (right) panel.
Lower row: Example for the contours $\Gamma_{\rm sef}$ in the complex $\sqrt{\sigma}$ plane. The pink and green lines in left (right) panel correspond to the integral contours for $D^0\pi^+$ and $D^+\pi^0$ ($D^0\pi^0$ and $D^+\pi^-$) self-energies, respectively.}
\label{fig:Dstp}
\end{center}
\end{figure*}
The self-energy on physical Riemann sheet in the upper half-plane of $\sqrt{\sigma}$ is connected
along the real axis, $\sqrt{\sigma}\in [m_{i,3}+m_{i,4},\infty)$, to the unphysical Riemann sheet in the lower half-plane. 
For the energy ${\rm Im }\sqrt{\sigma}<0$ (lower right in Fig.~\ref{fig:TBCont}), the self-energy on unphysical Riemann sheet can be obtained by deforming the contour $\Gamma_{\rm sef}$ as shown to the lower left in Fig.~\ref{fig:TBCont}. In particular, the singularity at $l_{cm}$ also acquires a negative imaginary part, but a smooth deformation of $\Gamma_{R}$ ensures that the contour $\Gamma_{\rm sef}$ still passes the singularity to the right. 
This guarantees that the self-energy has been analytically continued
from physical Riemann sheet to the unphysical Riemann sheet in the lower half-plane, where the resonance resides.

To display the contour $\Gamma_{\rm sef}$ and the pole position of $D^{*+}$ ($D^{*0}$) resonance 
evaluated from Eq.~\eqref{Eq:TwoSub} in the same plot, the contour $\Gamma_{\rm sef}$ is mapped to the $\sqrt{\sigma}$ plane according to $\sqrt{\sigma}=[\omega_{i,j}(l)+\omega_{i,k}(l)]^{1/2}$. In Fig.~\ref{fig:Dstp}, the pink and green lines in lower left panel (lower right panel) correspond to the integral contours for $D^0\pi^+$ and $D^+\pi^0$ ($D^0\pi^0$ and $D^+\pi^-$) self-energies, respectively. The pole position of $D^{*+}$ ($D^{*0}$) at $\sqrt{\sigma}$ plane is shown in the upper left panel (upper right panel) of Fig.~\ref{fig:Dstp}. One can see as the $D^0\pi^+$ and $D^+\pi^0$ channels are on their unphysical Riemann sheets, the pole position of $D^{*+}$ evaluated from Eq.~\eqref{Eq:TwoSub} is at 
\begin{eqnarray}
\label{eq:dspole}
{\rm Re}\,m_{D^{*+}}^{p}=m_{D^{*+}},\quad  {\rm Im}\,m_{D^{*+}}^{p}=-43.5 \,\rm keV. 
\end{eqnarray}
As the $D^0\pi^0$ and $D^+\pi^-$ channels are on their unphysical Riemann sheets, the pole position of $D^{*0}$
evaluated from Eq.~\eqref{Eq:TwoSub} is at 
\begin{eqnarray}
\label{eq:dzpole}
{\rm Re}\,m_{D^{*0}}^{p}=m_{D^{*0}}, \quad {\rm Im}\,m_{D^{*0}}^{p}=-32.5 \, \rm keV.
\end{eqnarray}
\begin{figure*}[tbhp]
\begin{center}
\includegraphics [scale=0.45] {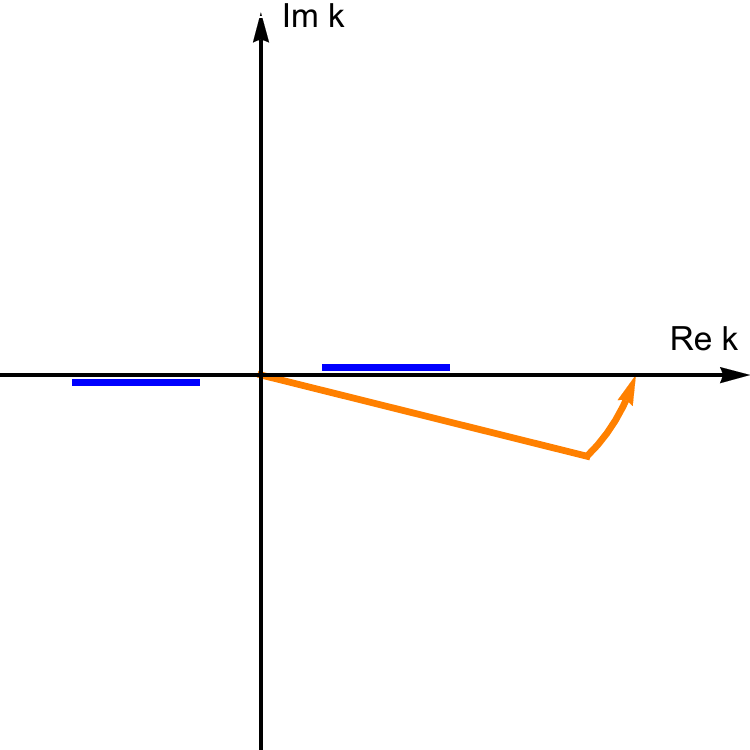}
\includegraphics [scale=0.45] {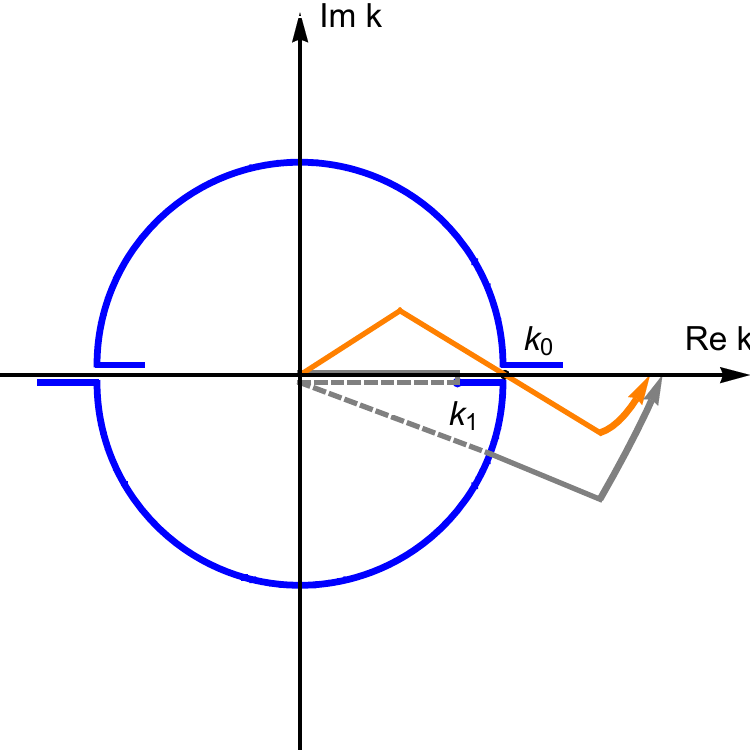}
\includegraphics [scale=0.45] {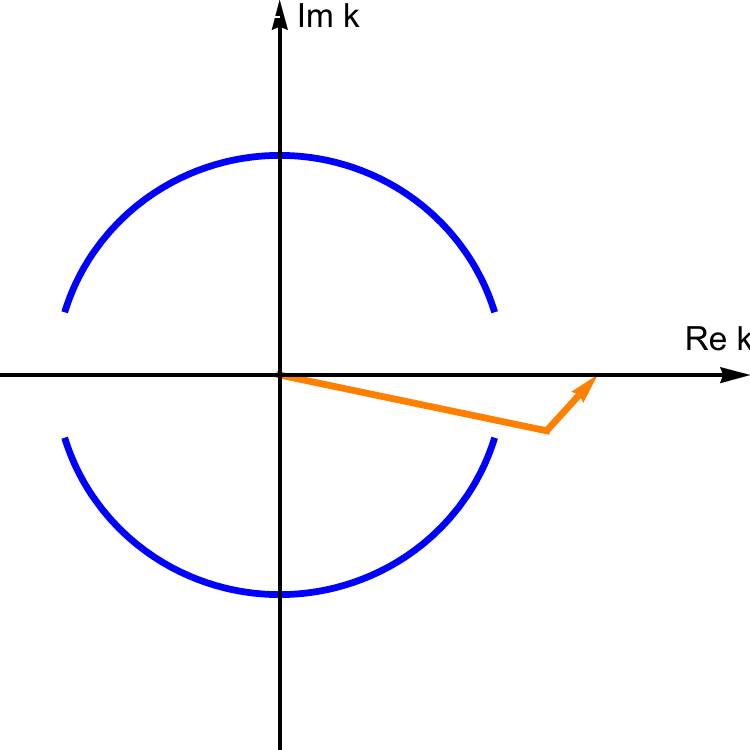}
\caption{The blue lines are positions of the singularities of $k$ for a fixed $p'$, $\sqrt{s}$ and $x\equiv {\rm cos} \theta \in [-1,1]$ for solution of $k_{\pm}$ in Eq.~\eqref{Eq:SinOpe}. In the medium panel, the momentum $k_0$ is the position where the contour is squeezed between the logarithmic cuts. The $k_1$ is the end point of the circular cut $k_-$ in 
Eq.~\eqref{Eq:SinOpe} by setting $x=+1$.}
\label{fig:opecut}
\end{center}
\end{figure*}

\begin{widetext}
\subsection{Singularities in the $\pi$-exchange potential}
\label{sec:SingPi}
Apart from singularities in two-body subsystem amplitude, there are singularities in the $\pi$-exchange potential. The singularity occurs when the denominator of Eq.~(\ref{Eq:OPE}) vanishes for any $x\equiv {\rm cos} \theta \in [-1,1]$ according to the partial wave decomposition of Eq.~\eqref{eq:patwa}. For a fixed three-body energy $\sqrt{s}$ and 
outgoing spectator momentum $p'$, the singularities are given by
\begin{eqnarray}
\label{Eq:SinOpe}
k_{\pm}=\frac{2\alpha p'x\pm \sqrt{4\alpha^2p'^2x^2+(\beta^2-4p'^2x^2)(\alpha^2-\beta^2m_{i,2}^2)}}{(\beta^2-4p'^2x^2)},
\end{eqnarray}
where
\begin{eqnarray}
\alpha&=&m_{\pi}^2- s-m_{i,2}^2-m_{i',2}^2+2\sqrt{s}\, \omega_{i',2}(p'), \nonumber \\
\beta&=&2\omega_{i',2}(p')-2\sqrt{s},
\end{eqnarray}
and $m_{\pi}$ is the exchanged $\pi^{0}$ or $\pi^{\pm}$ mass. In Fig.~\ref{fig:opecut}, as an example we show the positions of the singularities of $k$ for a fixed $p'$, $\sqrt{s}$ and $x\equiv {\rm cos} \theta \in [-1,1]$ for solution of $k_{\pm}$ in Eq.~\eqref{Eq:SinOpe}. In the complex $k$ plane,
as $0<p'<p_{1}'$ (left panel in Fig.~\ref{fig:opecut}), the singularities are located in the first and third quadrants. Increasing the value of $p'$, as $p_1'<p'<p_{2}'$ (medium panel in Fig.~\ref{fig:opecut}), a circular cut arises from the $\pi$-exchange potential. The circular cut grows with the increase of the value of $p'$. Still increasing the value of $p'$, as $p_{2}'<p'$ (right panel in Fig.~\ref{fig:opecut}), the circular cut opens. Here $p_{1}'$ and $p_{2}'$ are 
\begin{eqnarray}
p_{1}'=\frac{\lambda^{1/2}\Big((\sqrt{s}-m_{i,2})^2,m_{i',2}^2,m_{\pi}^2\Big)}{2(\sqrt{s}-m_{i,2})},\quad \quad  
p_{2}'=\frac{\lambda^{1/2}\Big(s,(m_{\pi}+m_{i,2})^2,m_{i',2}^2\Big)}{2\sqrt{s}}.
\end{eqnarray}
\end{widetext}

\subsection{Continuation across the three-body and complex two-body cuts}
\label{sec:AnaThree}
To explore the region of the complex $\sqrt{s}$ plane where the $T_{cc}^+$ is located, the amplitude $T(s,p',p)$ can be analytically continued into the unphysical Riemann sheets via the deformation of the integration contour. We make the transformation~\cite{Glockle:1978zz,Pearce:1984ca,Afnan:1991kb}
\begin{eqnarray}
\label{eq:RoTaAngular}
p' \to p'e^{-i\theta},\quad k \to ke^{-i\theta}, 
\end{eqnarray}
with $0<\theta<\frac{\pi}{2}$. As shown in the left panel in Fig.~\ref{fig:opecut}, the integration contour can be chosen along the orange line.
Since both $p'$ and $k$ in 
Eq.~\eqref{eq:bse} are rotated by a same angle $\theta$, the potential $V(s,p',k)$ does not create any problem in rotating the contour of integration. And we only need
to examine how the singularities from $\tau^i(\sigma_k)$ effect the contour rotation.

\begin{figure}[tbhp]
\begin{center}
\includegraphics [scale=0.35] {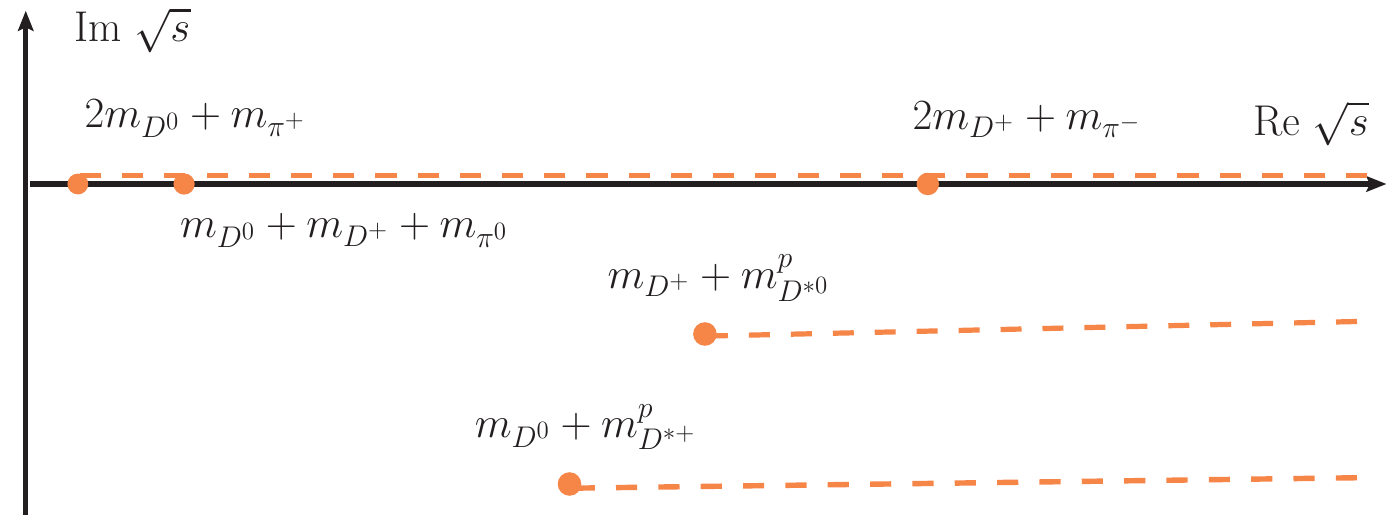}
\includegraphics [scale=0.27] {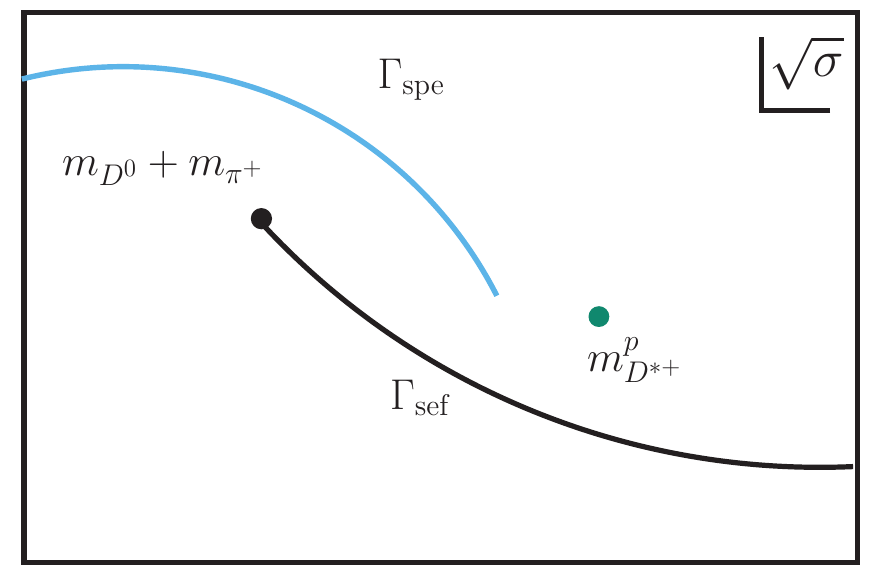}\quad
\includegraphics [scale=0.27] {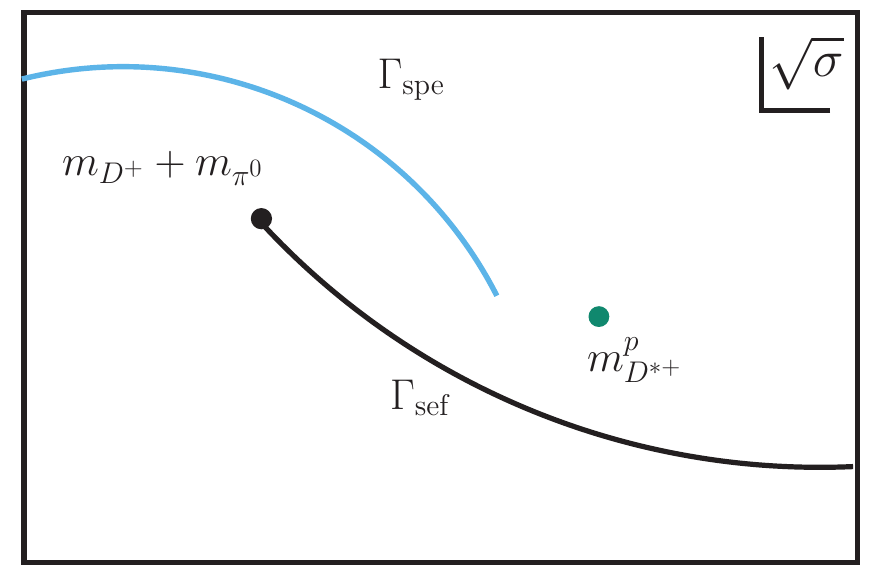}
\includegraphics [scale=0.27] {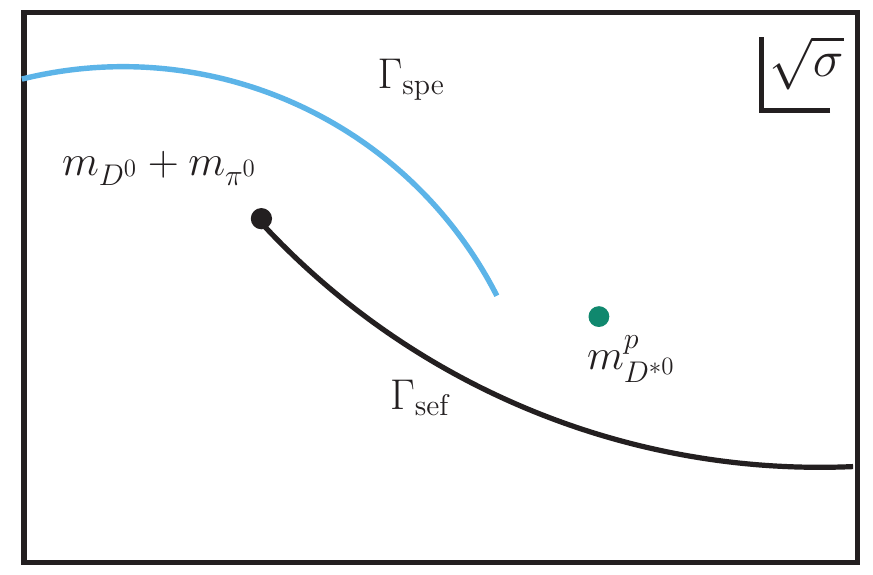}\quad
\includegraphics [scale=0.27] {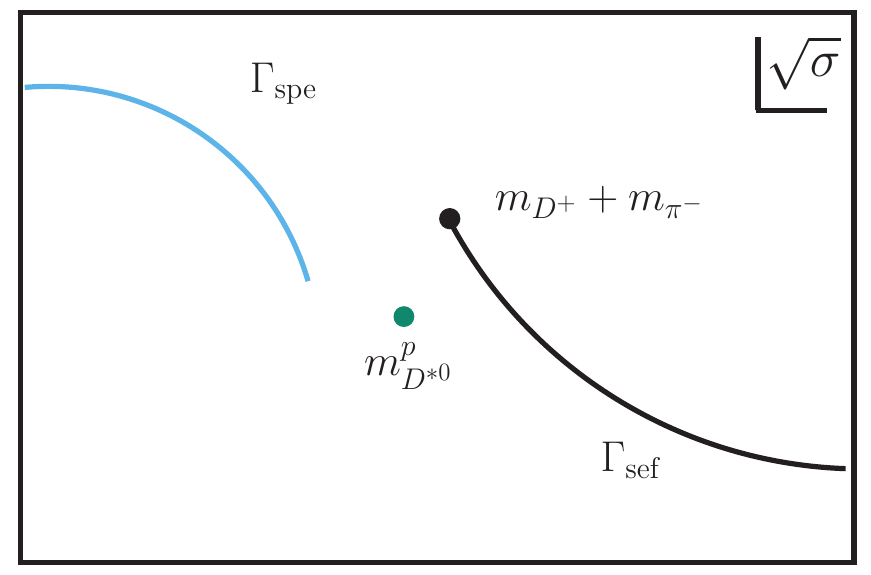}
\caption{Analytic structure of the three-body amplitude in the $\sqrt{s}$ plane. The real and complex branch
points (orange dots) are shown together with their respective cuts (orange dashed lines). The inserts show the contours $\Gamma_{\rm spe}$ mapped to the $\sqrt{\sigma}$ plane (blue lines), in a qualitative way. In the $\sqrt\sigma$ plane, the contours $\Gamma_{\rm sef}$ (black lines) starting at the two-body thresholds (black dots) do not change if $\sqrt{s}$ changes, but the contours $\Gamma_{\rm spe}$ do.
}
\label{fig:InteCont}
\end{center}
\end{figure}

The $D^0\pi^+$ ($D^+\pi^0$, $D^0\pi^0$ and $D^+\pi^-$) self-energy in Eq.~\eqref{eq:sefbar} has a well-known right-hand
cut along the real $\sqrt{\sigma}$ axis as discussed in \ref{sec:AnaSef}. It is clear that this induces a three-body cut in the propagator $\tau^i(\sigma_k)$ in three-body system which begins at $2m_{D^0}+m_{\pi^+}$ ($m_{D^0}+m_{D^+}+m_{\pi^0}$ and $2m_{D^+}m_{\pi^-}$). These three-body cuts along the real $\sqrt{s}$ axis are shown in Fig.~\ref{fig:InteCont}.
Here, we take the analytic continuation of the amplitude ${T}(s,p',p)$ across the three-body
$D^0D^0\pi^+$ cut as an example. By deforming the integration contour, the amplitude ${T}(s,p',p)$
can be analytically continued across the $D^0D^0\pi^+$ cut, for ${\rm Im}\sqrt{s}<0$ and ${\rm Re}\sqrt{s}>2m_{D^0}+m_{\pi^+}$. The inserts in Fig.~\ref{fig:InteCont} show the $\sqrt{\sigma}$ plane
with the contours $\Gamma_{\rm spe}$ and $\Gamma_{\rm sef}$. The position of the inserts in the $\sqrt{s}$ plane qualitatively corresponds to $\sqrt{s}$ used to map the contour $\Gamma_{\rm spe}$ to the $\sqrt{\sigma}$ plane, via Eq.~\eqref{eq:TwoInv}. Note the position of the contour $\Gamma_{\rm spe}$ relative to the $D^0\pi^+$ branch point at $m_{D^0}+m_{\pi^+}$ and $D^{*+}$ pole at $m_{D^{*+}}^p$ (upper left insert in Fig.~\ref{fig:InteCont}). Moreover, there is no across between the contours $\Gamma_{\rm spe}$ and $\Gamma_{\rm sef}$ in $\sqrt \sigma$ plane. This ensures that the integrand is always analytic. With the similar method, the amplitude ${T}(s,p',p)$ can be analytically continued across the three-body $D^0D^+\pi^0$ or $D^+D^+\pi^-$ cut.

Other singularities appear at the zero of the denominator of Eq.~\eqref{Eq:IsobarA}, 
\begin{eqnarray}
\label{eq:twobodydeno}
{\sigma_k-m_{i,1}^2-\Sigma_{i,34}^{R,{\rm \uppercase\expandafter {\romannumeral 2}}}(\sigma_k)-\Sigma_{i,3'4'}^{R,{\rm \uppercase\expandafter {\romannumeral 2}}}(\sigma_k)}+im_{i,1}\Gamma_{i,\gamma}=0.
\end{eqnarray}
Note that the self-energies are evaluated on their unphysical Riemann sheets. The pole positions of the resonances $D^{*+}$ and $D^{*0}$ are given by Eq.~\eqref{eq:dspole} and Eq.~\eqref{eq:dzpole}, respectively. Using $\sigma_{k}=\sqrt{s}-m_{i,2}$ at $k=0$, we obtain the positions of the complex branch points at complex $\sqrt{s}$ plane,
\begin{eqnarray}
\sqrt{s}=m_{i,1}^{p}+m_{i,2},
\end{eqnarray}
where $m_{i,1}^{p}$ is the pole position of $D^{*+}$ or $D^{*0}$.
Thus, the complex branch points in $\sqrt{s}$ plane are directly related to the poles of the unstable particles.

Translated into complex $k$ plane, the singularities corresponding to the vanishing of denominator of Eq.~\eqref{Eq:IsobarA} are at
\begin{eqnarray}
\label{eq:onshellk}
k_{b}= \frac{\lambda^{1/2}\big(s,(m_{i,1}^p)^2,m_{i,2}^2\big)}{2\sqrt{s}}.
\end{eqnarray}
For ${\rm Re}\sqrt{s}>m_{i,1}+m_{i,2}$, and ${\rm Im}\sqrt{s}< {\rm Im}\,m_{i,1}^p$,
those branch points are in the fourth quadrant of the $k$ plane and at an angle of $\phi_b$, where
\begin{eqnarray}
{\rm \tan} \phi_b=\frac{{\rm Im}\, k_b}{{\rm Re}\, k_b}.
\end{eqnarray}
Thus, as far as the complex branch point is concerned, we can take $\theta>\phi_b$ in Eq.~\eqref{eq:RoTaAngular} to analytically continue the amplitude ${T}(s,p',p)$ into another unphysical Riemann sheet. See Refs.~\cite{Pearce:1984ca,Afnan:1991kb} for a more detailed discussion.
\begin{figure}[tbhp]
\begin{center}
\includegraphics [scale=0.7] {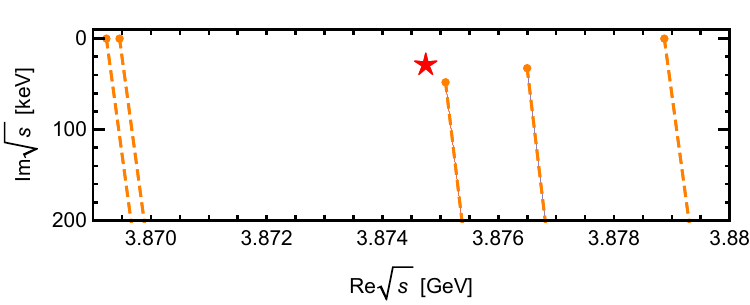}
\caption{The cut structure of the effective BS equation in complex $\sqrt{s}$ plane for the momentum contour rotated by an angle of $\theta=\pi/8$ in Eq.~\eqref{eq:RoTaAngular}. The real and complex branch
points (orange dots) are shown together with their respective cuts (orange dashed lines). The $T_{cc}^+$ pole observed  by the LHCb Collaboration at $\sqrt{s}$ plane is highlighted with the red star.}
\label{fig:TheetStr}
\end{center}
\end{figure}

In Fig.~\ref{fig:TheetStr}, we show the cut structure of the effective BS equation in complex $\sqrt{s}$ plane for the momentum contour rotated by an angle of $\theta=\pi/8$ in Eq.~\eqref{eq:RoTaAngular}. From Fig.~\ref{fig:TheetStr}, it is clear we have extended the energy domain over which ${T}(s,p',p)$ is defined to that part of the unphysical Riemann sheet where resonances are normally located.
As shown in the complex $\sqrt{s}$ plane, the $T_{cc}^+$ pole is always to the left of the complex branch points at $m_{D^{*+}}^{p}+m_{D^0}$ and $m_{D^{*0}}^{p}+m_{D^+}$, to the right of the branch points at $2m_{D^0}+m_{\pi^+}$ and $m_{D^0}+m_{D^+}+m_{\pi^0}$, and to the left of the branch point at $2m_{D^+}+m_{\pi^-}$. 
Therefore, the qualitative positions of $\Gamma_{\rm spe}$ and contours $\Gamma_{\rm sef}$ in the $\sqrt{\sigma}$ plane, corresponding  to $\sqrt{s}$ taking the value of $T_{cc}^+$ pole, are given by the inserts in Fig.~\ref{fig:InteCont}. The contours $\Gamma_{\rm spe}$ pass both the $D^{*+}$ and $D^{*0}$ poles to the left.
Both the $D^0\pi^+$ and $D^+\pi^0$ self-energy contours $\Gamma_{\rm sef}$ pass the $D^{*+}$ pole on the right.
The $D^0\pi^0$ and $D^+\pi^-$ self-energy contours $\Gamma_{\rm sef}$ pass the $D^{*0}$ pole on the right and left, respectively.

\subsection{The ${T}(s,p',p)$ for real momentum $p$ and $p'$}\label{sec:RealMom}
The effective BS equation is solved numerically by replacing the integrals by sums using Gaussian quadratures and inverting the resulting matrix equation~\cite{Haftel:1970zz}. To obtain ${T}(s,p',p)$ for real momentum $p$ and $p'$,
a major difficulty in doing this is the treatment of the singularities of the kernel. These arise both from the isobar propagator and the $\pi$-exchange potential. In the present work, we use the contour deformation method developed in Refs.~\cite{Aaron:1966zz,Cahill:1971ddy,BookThree}. In Ref.~\cite{Zhang:2023wdz}, some formal aspects of this technique were also described in detail. 
The method consists of two steps. Firstly, the analytically continuation of the effective BS equation for complex momentum is made. The effective BS equation is solved for complex momenta which is free from singularities. Then the solution for real momenta is obtained
using Cauchy’s theorem, starting with the solution for complex momenta.

Here, we would like to describe the detailed treatment of the singularities of the potential $V^{i'i}_{L'L}(s,p',p)$ having the configuration shown in Fig.~\ref{fig:opecut}.
For $0<p'<p_{1}'$ (left panel in Fig.~\ref{fig:opecut}), and for $p_2'<p'$ (right panel in Fig.~\ref{fig:opecut}), the integration contours can be chosen along the orange lines in the fourth quadrant of the complex
$k$ plane. For $p_1'<p'<p_2'$ (medium panel in Fig.~\ref{fig:opecut}), integration contour may be chosen along the gray line.
It is important that when integrating along this contour, one has to go onto the second sheet of $V^{i'i}_{L'L}(s,p',p)$ (dashed line part in the medium panel in Fig.~\ref{fig:opecut}). And the knowledge of the amplitude ${T}(s,p',p)$ in the interval $[0,k_1]$ is needed,
where $k_1=k_{-}(x=+1)$ in Eq.~(\ref{Eq:SinOpe}). However, one can
find that the inequality $p_1'>k_1$ does not always hold in this work.
In the present work, for $p_1'<p'<p_2'$, we use another integration contour (the orange line in the medium panel in Fig.~\ref{fig:opecut}). This contour has been  
employed in ~\cite{Cahill:1971ddy,Fix:2019txp}.
The potentials $V^{i'i}_{L'L}(s,p',p)$ are always calculated on the physical Riemann sheet. A certain disadvantage of this method lies in the fact that the position of the momentum $k_0$, where the contour is squeezed between the logarithmic cuts, depends on the value of the momentum $p'$. For this reason one has to solve the set of equation separately for each value of $p'$ of the chosen mesh.
\begin{widetext}
\begin{figure*}[tbhp]
\begin{center}
\includegraphics [scale=0.7] {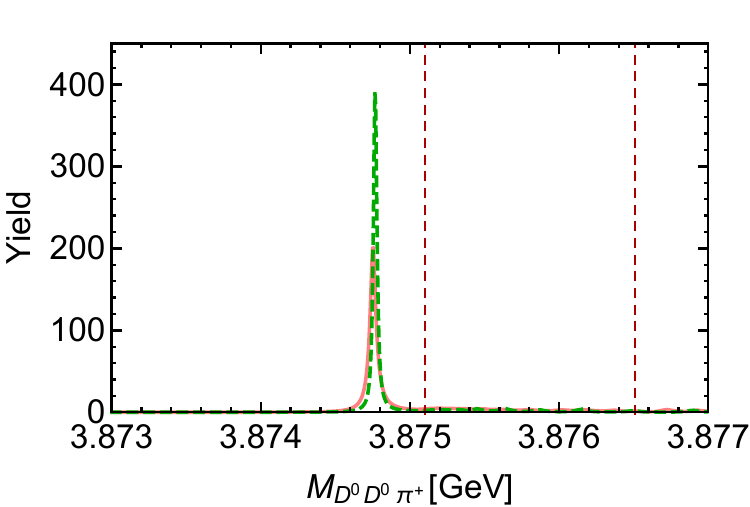}\quad
\includegraphics [scale=0.68] {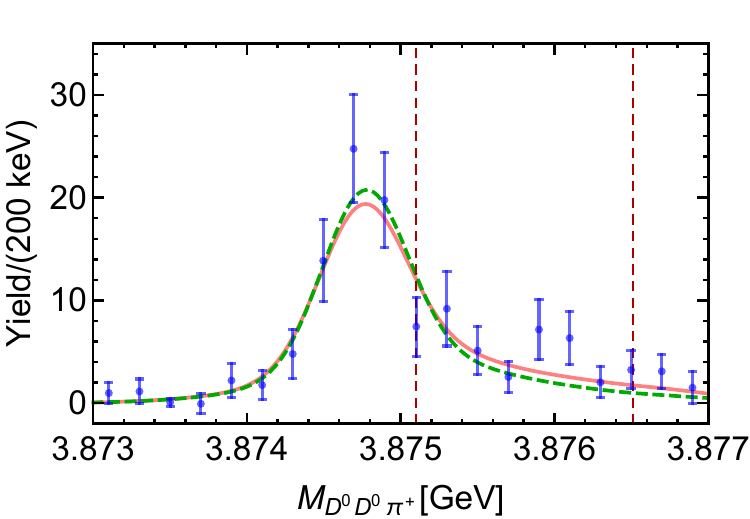}
\caption{Fitting results of the $D^0D^0\pi^+$ line shapes before (left panel) and after (right panel) convolution with the energy resolution function. The blue dashed and pink lines correspond to 
Schemes ${\rm \uppercase\expandafter {\romannumeral 1}}$ and ${\rm \uppercase\expandafter {\romannumeral 2}}$, respectively.
The vertical dashed lines represent the $D^0D^{*+}$ and $D^+D^{*0}$ thresholds. The experimental binning with the bin size of 200 keV is included in the fits.}.
\label{fig:FitRet}
\end{center}
\end{figure*}
\begin{table*}[tb]
\caption{The values of the parameters $\Lambda$ and the pole positions from fitting of the $D^0D^0\pi^+$ line shape
obtained by the LHCb Collaboration. The pole positions are given relative to the $D^0D^{*+}$ threshold.
The uncertainties of the cutoff $\Lambda$ are obtained by $\chi^2$ fitting to the LHCb data and propagate to the pole position.} 
\begin{ruledtabular}
\setlength{\tabcolsep}{5mm}{
\begin{tabular}{lcccc}
Scheme &  $\chi^2$/d.o.f  & $\Lambda$ GeV & $\sqrt{s_{pole}^{thr}}$ keV  \tabularnewline
\hline 
${\rm \uppercase\expandafter {\romannumeral 1}}$ &  $18.11/(20-1)=0.95$ & $0.4551\pm 0.0018$ & $-332^{+37}_{-36}-i(18\pm 1)$  \tabularnewline
${\rm \uppercase\expandafter {\romannumeral 2}}$&  $14.47/(20-1)=0.76$ & $0.3701\pm 0.0017$ & $-351^{+37}_{-35}-i(28\pm 1)$ 
\end{tabular}}
\end{ruledtabular}
\label{Tab:FitRes}
\end{table*}
\begin{table*}[tb]
\caption{The effective coupling constants extracted as defined in Eq.~(\ref{eq:EfeC}).} 
\begin{ruledtabular}
\setlength{\tabcolsep}{5mm}{
\begin{tabular}{lcccc}
Scheme &  $g^1$  & $g^2$ &  $g^{I=0}$ &  $g^{I=1}$  \tabularnewline
\hline 
${\rm \uppercase\expandafter {\romannumeral 1}}$ &  $3.90_{-0.09}^{+0.09}-i0.04_{-0.00}^{+0.00} $ & $-4.11_{-0.09}^{+0.09}+i0.04_{-0.00}^{+0.00} $ & $-5.66_{-0.13}^{+0.13}+i0.06_{-0.00}^{+0.00} $ & $0.15_{-0.00}^{+0.00}+i0.00_{-0.00}^{+0.00}$    \tabularnewline
${\rm \uppercase\expandafter {\romannumeral 2}}$&   $4.00_{-0.09}^{+0.09}+i0.04_{-0.00}^{+0.00} $ & $-4.13_{-0.09}^{+0.09}+i0.05_{-0.00}^{+0.00} $ & $-5.75_{-0.13}^{+0.13}+i0.01_{-0.00}^{+0.00} $
& $0.09_{-0.00}^{+0.00}-i0.07_{-0.00}^{+0.00}$ 
\end{tabular}}
\end{ruledtabular}
\label{Tab:EffeCo}
\end{table*}
\end{widetext}
\section{Numerical Results and Discussion}\label{sec:Results}
In the following we present the results of the approach in terms of the line shape
$\frac{d\Gamma(\sqrt{s})}{d\sqrt{s}}$ calculated from Eq.~(\ref{eq:lineshape}).
The free parameters of the model are the cutoff $\Lambda$ and the overall normalization factor $\mathcal{F}$. The two parameters are fixed by a fitting to the $D^0D^0\pi^+$ line shape obtained by the LHCb Collaboration~\cite{LHCb:2021vvq,LHCb:2021auc}. When fitting the LHCb data, the coupling constants $g_{L}^{i}$ can be absorbed into the overall factor $\mathcal{F}$. 
To take into account the experimental resolution, the line shape
$\frac{d\Gamma(\sqrt{s})}{d\sqrt{s}}$ in Eq.~(\ref{eq:lineshape}) is convolved with an energy resolution function, 
\begin{eqnarray}
\frac{d\widetilde{\Gamma}(\sqrt{s})}{d\sqrt{s}}=\int d\sqrt{s'} \, R_{\rm LHCb} (\sqrt{s},\sqrt{s'})\frac{d\Gamma(\sqrt{s'})}{d\sqrt{s'}}. 
\end{eqnarray}
The resolution function $R_{\rm LHCb} (\sqrt{s},\sqrt{s'})$ is parameterized by a sum of two Gaussian functions,
\begin{eqnarray}
 \nonumber R_{\rm LHCb} (\sqrt{s},\sqrt{s'})=\sum_{i=1,2}\alpha_i \frac{1}{\sqrt{2\pi}\sigma_i}{\rm exp}\Big(-\frac{(\sqrt{s'}-\sqrt{s})^2}{2\sigma_i^2}\Big), \\
\end{eqnarray}
where $\alpha_1=0.778$, $\alpha_2=0.222$, $\sigma_1=1.05\times 263$ keV,  $\sigma_2=2.413 \sigma_1$~\cite{LHCb:2021auc}.

In order to investigate the role played by the $\pi$-exchange potential, we consider the following
two different fit schemes:

(i) Scheme ${\rm \uppercase\expandafter {\romannumeral 1}}$: The OBE potentials excluding $\pi$-exchange potential.

(ii) Scheme ${\rm \uppercase\expandafter {\romannumeral 2}}$: The full dynamical calculation, the $\pi$-exchange potential is included.

Once the cutoff $\Lambda$ is fixed through the fit to the experimental data, 
the position of the pole of the amplitude can be searched in the complex energy
$\sqrt{s}$ plane. The physical $T_{cc}^+$ signal is associated with the corresponding pole in the $D^0D^{*+}-D^+D^{*0}$ scattering amplitude ${T}(s,p',p)$.
The effective coupling $g^i$ to channel $i$ can be obtained from the
residue of the amplitude ${T}(s,p',p)$ at the pole position,
\begin{eqnarray}
\label{eq:EfeC}
g^{i'}g^{i}= \mathop{\rm lim}_{s\to s_{pole}}\frac{1}{4\pi} (s-s_{pole})T^{i'i}(s,k_b,k_b),
\end{eqnarray}
and the on shell momentum $k_b$ is given in Eq.~(\ref{eq:onshellk}). 

The fitted line shapes for the two schemes are shown in Fig.~\ref{fig:FitRet}. The parameters of the fits and extracted pole positions are given in Table~\ref{Tab:FitRes}.
These fitting quantities can be assessed through the corresponding values of $\chi^2$/d.o.f given in Table~\ref{Tab:FitRes}. The fitting results of the two schemes are comparable.

The three-body and complex two-body cuts imply the multi-Riemann sheets of the amplitude
${T}(s,p',p)$. As shown in Fig.~\ref{fig:TheetStr}, in the energy near the $T_{cc}^+$ pole, the $D^0D^0\pi^+$ and $D^+D^0\pi^+$ channels are on their unphysical Riemann sheets, while the $D^+D^+\pi^-$ is on its physical Riemann sheet.
In Scheme ${\rm \uppercase\expandafter {\romannumeral 1}}$, where OBE potentials excluding $\pi$-exchange potential, the position of the pole is at $\sqrt{s_{pole}^{thr}}=-332^{+37}_{-36}-i(18\pm 1)$ keV. In Scheme ${\rm \uppercase\expandafter {\romannumeral 2}}$, where the $\pi$-exchange potential is included, the position of the pole is at $\sqrt{s_{pole}^{thr}}=-351^{+37}_{-35}-i(28\pm 1)$ keV. The width of $T_{cc}^+$
in Scheme ${\rm \uppercase\expandafter {\romannumeral 2}}$ is larger than 
in Scheme ${\rm \uppercase\expandafter {\romannumeral 1}}$ by a factor of 1.5. Our results indicate that the inclusion of $\pi$-exchange mainly influences the width of $T_{cc}^+$. For a bound state with a small binding energy, the $\pi$-exchange can be dealt with using perturbation theory. This is agreement with the findings in Refs.~\cite{Fleming:2007rp,Baru:2011rs,Schmidt:2018vvl,Braaten:2015tga}.

In Table~\ref{Tab:EffeCo}, we show the values of the effective coupling constants to different channels obtained from Eq.~(\ref{eq:EfeC}). In the exact isospin limit, one would have $g_{L}^1=-g_{L}^2$ ($g_{L}^1=g_{L}^2$) for an isoscalar (isovector) state.
We find that the coupling constants $g_{L}^1$ and $g_{L}^2$ are very 
close to each other with an opposite sign. This indicates that
we have basically a state with an isospin $I=0$. The values of the coupling constants obtained here are similar to those obtained in Refs.~\cite{Feijoo:2021ppq,Albaladejo:2021vln,Du:2021zzh}.

\section{Summary}\label{sec:Sum}
In this work, we discussed the analytic continuation of the three-body $D^0D^{*+}-D^+D^{*0}$ scattering amplitude ${T}(s,p',p)$. Compared with the two-particle scattering, complications arise for the three-particle scattering not only because of the increase in the number of variables necessary to describe the processes, but also the possible appearance of the
dynamic $\pi$-exchange, three-body and complex two-body unitarity cuts. In particular, we find that the logarithmic singularities of the $\pi$-exchange potential can form into a circular cut. Via the contour
deformation, this cut can be circumvented, and the integration of the effective BS equation does not pose any numerical problem.

Employing the contour deformation, the effective BS equation can be analytically 
continued to the unphysical region. As we have shown, one can choose a self-consistent integration contour, which defines a smooth continuation of the amplitude to 
the domain of analyticity. As an implementation, we find the $\pi$-exchange term has a signification on the pole position of the $T_{cc}^+$. Including the $\pi$-exchange term, the width of $T_{cc}^+$ will be increased by a factor of 1.5.

Systematic analysis the newly observed hadronic states requires building the amplitude that satisfies the constrains such as unitarity and analyticity. The present work discussed the prescription for solving and analytically continuing the effective BS equation describing the three-body reactions. Such an analysis is expected to provide an important theoretical background for determining the parameters of the newly exotic candidates. In the near future, we will extend our framework to calculate the $3\pi$-$K\bar K\pi$ coupled system suggested to be responsible for the exotic candidate $a_1(1420)$~\cite{COMPASS:2015kdx,Mikhasenko:2015oxp,Aceti:2016yeb,COMPASS:2020yhb}.

\acknowledgments 
I would like to thank Feng-Kun Guo for fruitful and enlightening discussions and comments. Also I would like to thank Jia-Jun Wu for useful discussions regarding the circular cut.
The results described in this paper are supported by HPC Cluster of ITP-CAS. This work is supported in part by the National Natural Science Foundation of China (NSFC) under Grants No. 12247139, by the Chinese Academy of Sciences under Grants No. XDB34030000 and No. YSBR-101; by the National Key R\&D Program of China under Grant No. 2023YFA1606703.

\appendix
\begin{widetext}
\section{The interaction Lagrangian}\label{sec:APPLa}
The interaction Lagrangian between pseudo-Goldstone bosons and the mesons containing a heavy quark can be 
constructed by imposing invariance under both heavy quark spin-flavor transformation and chiral transformation~\cite{Burdman:1992gh,Wise:1992hn,Yan:1992gz,Falk:1992cx,Casalbuoni:1996pg}. The light vector mesons nonet can be introduced by using the hidden
gauge symmetry approach~\cite{Casalbuoni:1992dx,Casalbuoni:1992gi,Casalbuoni:1996pg}. The Lagrangian containing these particles can be written as
\begin{eqnarray}
\nonumber&&{\cal L}_{ DD^{*}P}=g_{ DD^{*}P}({
D}_b { D}^{*\mu\dagger}_a+{ D}^{*\mu}_{b}{\
D}^{\dagger}_a) (\partial_{\mu}{\cal M})_{ba} +g_{\bar{D}\,\bar{D}^{*}P}(\,\bar{
D}^{*\mu\dagger}_a\bar{
D}_b+\bar{
D}^{\dagger}_a\bar{
D}^{*\mu}_b)(\partial_{\mu}{\cal M})_{ab},\\
\nonumber&&{\cal L}_{{\rm DD}V}= ig_{{ DD}V}({
D}_b\stackrel{\leftrightarrow}{\partial_{\mu}}{
D}^{\dagger}_a)V^{\mu}_{ba}+ig_{\bar{ D}\,\bar{
D}V}(\bar{
D}_b\stackrel{\leftrightarrow}{\partial_{\mu}}\bar{
D}^{\dagger}_a)V^{\mu}_{ab},\\
\nonumber &&{\cal L}_{{D^{*}D^{*}}V}= ig_{{ D^{*}D^{*}}V}({
D}^{*}_{b\nu}\stackrel{\leftrightarrow}{\partial_{\mu}}{
D}^{*\nu\dagger}_a)V^{\mu}_{ba}+ig'_{{ D^{*}D^{*}}V}({
D}^{*\mu}_{b}{ D}^{*\nu\dagger}_{a}-{ D}^{*\mu\dagger}_{a}{
D}^{*\nu}_{b})(\partial_{\mu}V_{\nu}-\partial_{\nu}V_{\mu})_{ba}\\
\nonumber&&\qquad \qquad+ ig_{\bar{D}^{*}\bar{
D}^{*}V}(\bar{
D}^{*}_{b\nu}\stackrel{\leftrightarrow}{\partial_{\mu}}\bar{ D}^{*\nu\dagger}_{a})V^{\mu}_{ab}+ig'_{\bar{D}^{*}\bar{D}^{*}V}(\bar{D}^{*\mu}_b\bar{D}^{*\nu\dagger}_a-\bar{D}^{*\mu\dagger}_a\bar{ D}^{*\nu}_b)(\partial_{\mu}V_{\nu}-\partial_{\nu}V_{\mu})_{ab},\\
&&{\cal L}_{D^*DV} =
  ig_{ D^{*}DV}\varepsilon_{\lambda\alpha\beta\mu}
  ( D_b {\stackrel{\leftrightarrow \lambda}{\partial}} D^{*\mu\dag}_a
  +D_b^{*\mu}{\stackrel{\leftrightarrow \lambda}{\partial}} D^\dag_a )
  (\partial^\alpha{}V^\beta)_{ba}+
  ig_{ {\bar D}^{*}\bar{D}V}\varepsilon_{\lambda\alpha\beta\mu}
  ( \bar{D}_b {\stackrel{\leftrightarrow \lambda}{\partial}} \bar{D}^{*\mu\dag}_a
  +\bar{D}_b^{*\mu}{\stackrel{\leftrightarrow \lambda}{\partial}} \bar{D}^\dag_a )
  (\partial^\alpha{}V^\beta)_{ba}. \quad \quad
\end{eqnarray}
The matrix ${\cal M}$ contains
$\pi$, $K$, $\eta$ fields, which is  a $3\times 3$ hermitian and
traceless matrix. $V_{\mu}$ is analogous to ${\cal M}$
containing $\rho$, $K^{*}$, $\omega$ and $\phi$. The matrix ${\cal M}$ and $V_{\mu}$ are expressed as
\begin{equation}
\label{5}{\cal M}=\left(
\begin{array}{ccc}
\frac{\pi^0}{\sqrt{2}}+\frac{\eta}{\sqrt{6}}&\pi^{+}&K^{+}\\
\pi^{-}&-\frac{\pi^{0}}{\sqrt{2}}+\frac{\eta}{\sqrt{6}}&K^{0}\\
K^{-}&\overline{K}^{0}&-\sqrt{\frac{2}{3}}\,\eta
\end{array}
\right), \quad 
V=\left(\begin{array}{ccc}
\frac{\rho^0}{\sqrt{2}}+\frac{\omega}{\sqrt{2}}&\rho^{+}&K^{*+}\\
\rho^{-}&-\frac{\rho^0}{\sqrt{2}}+\frac{\omega}{\sqrt{2}}&K^{*0}\\
K^{*-}&\overline{K}^{*0}&\phi
\end{array}
\right).
\end{equation}
The isospin doublets are $D=(D^+,-D^0)$ and $D^*=(D^{*+},-D^{*0})$.
The coupling constants are as follows,
\begin{eqnarray}
&& g_{
\nonumber DD^{*}P}=-g_{\bar{D}\,\bar{D}^{*}P}=-\frac{2g}{f_{\pi}}\sqrt{
M_{ D}M_{ D^{*}}}, \quad 
g_{{ DD}V}=-g_{\bar{ D}\,\bar{
D}V}=\frac{1}{\sqrt{2}}\beta g_{V},\quad \quad   g_{{ D^{*}D^{*}}V}=-g_{{\bar{
D}^{*}\bar{ D}^{*}}V}=-\frac{1}{\sqrt{2}}\,\beta g_{V}, \\
&&g'_{{ D^{*}D^{*}}V}=-g'_{{\bar{
D}^{*}\bar{ D}^{*}}V}=-\sqrt{2}\,\lambda g_{V}{
M_{D^{*}}}, \quad \quad  g_{{ D D^{*}}V}=g_{{ \bar{D} \bar{D}^{*}}V}=\sqrt{2}\,\lambda g_{V}.
\end{eqnarray}
The effective Lagrangian between
$\sigma$ and heavy mesons are 
\begin{eqnarray}
{\cal L}_{DD\sigma}=g_{ DD\sigma}\,{ D}_a{
D}^{\dagger}_a\sigma+g_{\bar{D}\,\bar{D}\sigma}\,\bar{
D}_a\bar{ D}^{\dagger}_a\sigma, \quad {\cal L}_{ {D}^{*}{D}^{*}{\sigma}}=g_{
D^{*}D^{*}\sigma}\, { D}^{*\mu}_a{ D}^{*\dagger}_{a\mu}\sigma+
g_{\bar{D}^{\,*}\bar{D}^{\,*}\sigma}\,
\bar{ D}^{*\mu}_a\bar{D}^{*\dagger}_{a\mu}\sigma,
\end{eqnarray}
and the relevant coupling constants are
\begin{eqnarray}
\nonumber&&g_{DD\sigma}=g_{
\bar{D}\,\bar{D}\,\sigma}=-2g_{\sigma}{M_D},  \quad \quad g_{D^{*}D^{*}\sigma}=g_{\bar{D}^{*}\bar{D}^{*}\sigma}=2g_{\sigma}{M_{D^{*}}}.
\end{eqnarray}
In this work, we choose the coupling constants $g=0.59$, $g_V=5.8$, $\beta=0.9$, $\lambda=0.56~\rm{GeV}^{-1}$, $f_{\pi}=0.132$~GeV and $g_{\sigma}=0.76$ as in Refs.~\cite{Falk:1992cx,Isola:2003fh}.

\section{The polarization vectors and kinematics}\label{sec:Polari}
Spin-1 helicity polarization vectors are given by
\begin{align}
\label{eq:}
\epsilon_0^{\mu}(p)=\frac{1}{m}\begin{pmatrix} |\vec{p}\,| \\  E \sin{\theta}\cos{\phi} \\ E \sin{\theta}\sin{\phi} \\ E \cos{\theta} \end{pmatrix}, \quad \quad
\epsilon_{\pm 1}^{\mu}(p)=\frac{1}{\sqrt{2}}\begin{pmatrix} 0 \\  \mp \cos{\theta}\cos{\phi}+i\sin{\phi} \\ \mp \cos{\theta}\sin{\phi}-i\cos{\phi} \\ \pm \sin{\theta} 
   \end{pmatrix},
\end{align}
where $p^{\mu}=(\,E,\,\vec{p}\,)$ is the particle four-momentum, $m$ is the particle mass and 
\begin{align}
\vec{p}=(\,|\vec{p}\,|\, {\rm sin} \theta \, {\rm cos}\phi,\, |\vec{p}\,|\, {\rm sin} \theta\, {\rm sin}\phi, |\vec{p}\,|\, {\rm cos}\theta\,).
\end{align}
The helicity sum gives
\begin{eqnarray}
\sum_{\lambda}\epsilon_{\lambda,\mu}(p) \epsilon_{\lambda,\nu}^{*}(p)=-g_{\mu\nu}+\frac{p_{\mu} p_{\nu}}{m^2}.
\end{eqnarray}
\end{widetext}
\bibliography{ref.bib}
 
\end{document}